\documentclass[10pt]{article}

\usepackage{amsmath}
\usepackage{amsfonts}
\usepackage{amsthm}
\usepackage[english]{babel}
\usepackage[all]{xy}

\usepackage{rotating}
\usepackage{graphicx}
\usepackage{longtable}
\usepackage{tikz}

\usepackage{enumerate}

\newtheorem{theorem}{Theorem}[section]

\newtheorem{lemma}[theorem]{Lemma}
\newtheorem{proposition}[theorem]{Proposition}

\newtheorem{definition}[theorem]{Definition}
\newtheorem{remark}[theorem]{Remark}

\theoremstyle{remark}

\newcommand{\mR}{\mathbb{R}}
\newcommand{\mC}{\mathbb{C}}
\newcommand{\mN}{\mathbb{N}}
\newcommand{\mE}{\mathbb{E}}
\newcommand{\mZ}{\mathbb{Z}}
\newcommand{\mS}{\mathbb{S}}
\newcommand{\mP}{\mathbb{P}}

\newcommand{\mH}{\mathbb{H}}

\newcommand{\cD}{\mathcal{D}}

\newcommand{\cM}{\mathcal{M}}
\newcommand{\cH}{\mathcal{H}}

\newcommand{\cF}{\mathcal{F}}
\newcommand{\cP}{\mathcal{P}}
\newcommand{\cC}{\mathcal{C}}

\newcommand{\cS}{\mathcal{S}}
\newcommand{\cG}{\mathcal{G}}
\newcommand{\cL}{\mathcal{L}}

\newcommand{\cJ}{\mathcal{J}}
\newcommand{\cI}{\mathcal{I}}
\newcommand{\cU}{\mathcal{U}}
\newcommand{\cT}{\mathcal{T}}

\newcommand{\dD}{\textbf{D}}

\newcommand{\bx}{{\bf x}}
\newcommand{\by}{{\bf y}}

 \def\a{{\alpha}} 
 \def\b{{\beta}}
 
 \def\k{{\kappa}}
 
 \def\l{{\lambda}}
 \def\d{{\delta}}
 \def\o{{\omega}}
 \def\s{{\sigma}}
 \def\la{{\langle}}
 \def\ra{{\rangle}} 

\newcommand{\ux}{\underline{x}}

\newcommand{\uy}{\underline{y}}

\newcommand{\uA}{\underline{\alpha}}

\newcommand{\upx}{\partial_{\underline{x}}}
\newcommand{\upy}{\partial_{\underline{y}}}

\begin{document}

\title{Clifford algebras, Fourier transforms and quantum mechanics}

\author{H. De Bie}

\date{\small{Department of Mathematical Analysis}\\
\small{Faculty of Engineering and Architecture -- Ghent University\\ Krijgslaan 281, 9000 Gent,
Belgium}}

\maketitle

\begin{abstract}

In this review,  an overview is given of several recent generalizations of the Fourier transform, related to either the Lie algebra $\mathfrak{sl}_{2}$ or the Lie superalgebra $\mathfrak{osp}(1|2)$. In the former case, one obtains scalar generalizations of the Fourier transform, including the fractional Fourier transform, the Dunkl transform, the radially deformed Fourier transform and the super Fourier transform. In the latter case, one has to use the framework of Clifford analysis and arrives at the Clifford-Fourier transform and the radially deformed hypercomplex Fourier transform. 

A detailed exposition of all these transforms is given, with emphasis on aspects such as eigenfunctions and spectrum of the transform, characterization of the integral kernel and connection with various special functions.
\end{abstract}

\maketitle

\tableofcontents

\section{Introduction}

Recently, the topic of generalized Fourier transforms related to realizations of the Lie algebra $\mathfrak{sl}_{2}$ or the Lie superalgebra $\mathfrak{osp}(1|2)$ has received considerable attention. In particular, the fractional Fourier transform, the Dunkl transform, the radially deformed Fourier transform and the super Fourier transform are scalar generalizations of the Fourier transform related to $\mathfrak{sl}_{2}$. The Clifford-Fourier transform and the radially deformed hypercomplex Fourier transform on the other hand are Clifford-algebra valued generalizations of the Fourier transform, related to realizations of $\mathfrak{osp}(1|2)$.

The present paper aims at providing an extensive review of the results that have been obtained for these generalized transforms. The emphasis will lie on aspects such as eigenfunctions and spectrum of the transforms under consideration, characterization of the integral kernel and connection with various special functions. As the number of publications related to all these transforms is huge, we cannot cover them all in the references. Rather, we restrict ourselves to those which are crucial for the exposition.

To focus the ideas that will be pursued in this review, let us revisit the classical Fourier transform (FT) in $\mR^{m}$. This transform can be defined in many ways. For us, 4 different formulations are in particular important. In its most basic formulation, the FT is of course given by the integral transform
\[
\textbf{F1}\quad  \cF(f) (y) = \frac{1}{(2 \pi)^{m/2}} \ \int_{\mathbb{R}^m} e^{-i \la x,y \ra} \ f(x) \ dx, \qquad f \in L^1(\mR^m)
\]
with $i$ the complex unit, $\la x,y \ra$ the standard inner product of $x, y \in \mR^m$ and $dx$ the Lebesgue measure on $\mathbb{R}^m$.
Alternatively, one can rewrite the transform as
\[
\textbf{F2}\quad \cF (f) (y) = \frac{1}{(2 \pi)^{m/2}} \ \int_{\mathbb{R}^m} K(x,y) \ f(x) \ dx
\]
where $K(x,y)$ is the unique solution of the system of PDEs
\[
\partial_{x_{j}} K(x,y) = - i y_{j} K(x,y), \quad j =1, \ldots, m
\]
under the inital condition $K(0,y)=1$. 
A third formulation is given by
\[
\textbf{F3}\quad \cF = e^{ \frac{i \pi m}{4}} e^{\frac{i \pi}{4}(\Delta - |x|^{2})}
\]
with $\Delta$ the Laplacian in $\mR^{m}$. This expression connects the Fourier transform with the Lie algebra $\mathfrak{sl}_{2}$ generated by $\Delta$ and $|x|^{2}$ and with the theory of the quantum harmonic oscillator, determined by the hamiltonian $H = -(\Delta - |x|^2)/2$.
Finally, the kernel can also be expressed as an infinite series in terms of special functions as (see \cite[Section 11.5]{MR0010746})
\[
\textbf{F4}\quad K(x, y) = 2^{\lambda} \Gamma(\lambda)\sum_{k=0}^{\infty}(k+ \lambda) (-i)^{k} (|x||y|)^{-\lambda} J_{k+ \lambda}(|x||y|) \; C_{k}^{\lambda}(\langle\xi,\eta \rangle),
\]
where $\xi= x/|x|$, $\eta = y/|y|$ and $\lambda =(m-2)/2$. Here, $J_{\nu}$ is the Bessel function and $C_{k}^{\l}$ the Gegenbauer polynomial.

Each formulation has its specific advantages and uses. The standard formulation $\textbf{F1}$ allows to immediately compute a bound of the kernel and is hence ideal to study the transform on $L^{1}$ spaces or more general function spaces. Formulation $\textbf{F2}$ yields the calculus properties of the transform, and allows to generalize the transform to e.g. the so-called Dunkl transform (see \cite{deJ} and Section \ref{DTF}).  Formulation $\textbf{F3}$ emphasizes the structural (Lie algebraic) properties of the Fourier transform and also allows to compute its eigenfunctions and spectrum.  Finally, $\textbf{F4}$ connects the Fourier transform with the theory of special functions, and is the ideal formulation to obtain e.g. the Bochner identities (which are a special case of the subsequent Proposition \ref{Bochner}).

As stated in the beginning, the aim of this review is to discuss several extensions of the classical FT. 
These can be divided into two classes: scalar transforms, which will be related to realizations of the Lie algebra $\mathfrak{sl}_2$, and hypercomplex transforms, related to realizations of  the Lie superalgebra $\mathfrak{osp}(1|2)$, which contains $\mathfrak{sl}_2$ as its even subalgebra. The latter transforms are defined using the language of Clifford analysis, developed in e.g. \cite{MR697564, MR1169463, GM}. For both classes of transforms, we will see that all four definitions \textbf{F1} - \textbf{F4} are necessary to obtain a complete description of the generalized Fourier transform under consideration.

Conceptually, the study of such generalized FTs is divided in several steps, which may be summarized by:
\begin{itemize}
\item find $\mathfrak{sl}_2$ or $\mathfrak{osp}(1|2)$ realization using differential and/or differential-difference operators
\item solve spectral/quantum problem for associated hamiltonian $H$
\item use exponential definition \textbf{F3} to define generalized FT
\item find series expansion of the form \textbf{F4}
\item find explicit expression of the form \textbf{F1}
\item study analytic properties of the obtained transform.
\end{itemize}

The paper is organized as follows. In Section 2 we discuss generalized scalar Fourier transforms, including the fractional Fourier transform, the Dunkl transform, the radially deformed Fourier transform and the super Fourier transform. In Section 3 we move to the subject of hypercomplex transforms. We start with a short exposition on Clifford algebras and analysis. Subsequently, we turn our attention to the so-called Clifford-Fourier transform, the hypercomplex transform which has received the most attention in the literature. Next, we consider the radially deformed hypercomplex Fourier transform, where we also give the connection with the Dunkl transform and the radially deformed Fourier transform. We also briefly discuss other hypercomplex transforms that have received considerable attention in the literature, but that do not fit into the framework described above. We end this paper by indicating a few open problems and directions for future research.

\section{Scalar Fourier transforms}
\label{Secsl}

Harmonic analysis in $\mR^{m}$ is governed by the following three operators
\begin{align*}
\Delta &:= \sum_{i=1}^{m}\partial_{x_{i}}^{2}\\
|x|^{2} &:=  \sum_{i=1}^{m}x_{i}^{2}\\
\mE &:= \sum_{i=1}^{m} x_{i} \partial_{x_{i}}
\end{align*}
with $\Delta$ the Laplace operator and $\mE$ the Euler operator. The operators $E = |x|^{2}/2$, $F =-\Delta/2$ and $H =\mE + m/2$ are invariant under $O(m)$ and generate the Lie algebra $\mathfrak{sl}_{2}$ (see e.g. \cite{MR1151617}):
\begin{equation}
\label{sl2relclass}
\big[H,E\big] = 2E,\>\> \big[H,F\big] = -2F,\>\>\big[E,F\big] = H.
\end{equation}
So there are two types of symmetries present: the orthogonal symmetry of the individual operators, as well as the Lie algebraic symmetry generated by the three operators together. In this section, we will consider four generalizations of the Fourier transform, each related to a generalized Laplacian that still leads to the same algebraic $\mathfrak{sl}_{2}$ symmetry, but has a group symmetry which may differ from the orthogonal group.

The fractional Fourier transform is a generalization of the FT that is still invariant under the orthogonal group $O(m)$. For the Dunkl transform, the symmetry is reduced to that of a finite reflection group $\cG < O(m)$. The radially deformed Fourier transform encompasses both the fractional and Dunkl transform. Finally, the super Fourier transform is defined in the context of superspaces and is invariant under the product of the orthogonal with the symplectic group.

For an overview of harmonic analysis related to these three different types of symmetry, we refer the reader to \cite{CPLMS}, in particular concerning the Hermite functions for each symmetry and their properties.

\subsection{The fractional Fourier transform}
\label{sec_frFT}

The fractional Fourier transform is a generalization of the classical Fourier transform (FT). It is usually defined using the operator expression 
\[
\cF_{\alpha} =    e^{ \frac{i \alpha m}{2}}e^{ \frac{i \alpha}{2}(\Delta - |x|^{2} )}, \quad \a \in [-\pi,\pi]
\]
with $\Delta$ the Laplace operator in $\mR^{m}$. As integral transform, it can be written as
\begin{align*}
\cF_{\alpha} (f)(y)&= \left(\pi (1- e^{-2i \alpha})\right)^{-m/2} \int_{\mR^m}  e^{-i \la x,y \ra / \sin \alpha} e^{\frac{i}{2}( \cot \alpha) (|x|^2 + |y|^2)}f(x) dx
\end{align*}
when $\a \not = \pm \pi$, $\a \not = 0$. The most important property of this transform is given by the semigroup identity
\[
\cF_{\alpha} \cF_{\beta} =\cF_{\alpha + \beta},
\]
thus explaining its name. Note that the ordinary FT corresponds to $\alpha = \pi/2$.

Historically, the fractional FT has been reinvented several times, sometimes even simultaneously by different scientific communities. In the applied literature, Namias \cite{MR573153} is often credited with its invention. However, Mustard \cite{MR1668139} attributes the fractional FT to Condon and Bargmann (see \cite{Con, Barg}). 
In physics, this transform was introduced independently in the seventies by Collins in optics \cite{Col}, as well as by Moshinsky and Quesne in theoretical physics \cite{MR0286385, MR0286386}. In the latter case, the fractional FT is part of a wider class of operators which were coined linear canonical transforms. In subsequent work \cite{MR0378646, MR0378644, MR0378645, MR0529766}, Wolf has further elaborated on this topic.

In pure mathematics, the main ideas leading to the discovery of the fractional FT seem to have been around implicitly since the discovery of the so-called Mehler formula \cite{M}, connecting the kernel of the fractional Fourier transform with a series expansion in terms of Hermite functions. In this context, one uses the term Hermite semigroup to denote the fractional FT, see e.g. \cite{MR0974332, MR0983366}. Note that in this situation the parameter $\alpha$ is usually extended to the right-half complex plane.
Even in the completely different mathematical field of $C^{*}$-algebras, a special case of the fractional FT has been introduced independently, see \cite{MR2055227}.

For a detailed overview of the theory and recent applications of the fractional FT we refer the reader to \cite{OZA}. A perspective from the point of view of the Hermite semigroup can be found in e.g. \cite{ST, Orsted2}.

Let us now state the eigenfunctions of this transform. For that aim, we introduce the Hermite basis for $L^2(\mR^m)$, resp. $\cS(\mR^m)$. The Hermite functions are given by
\begin{equation}
\label{HermFrac}
\phi_{j, k, \ell}:= L_{j}^{\frac{m}{2} + k-1}(|x|^2) H_k^{(\ell)} \, e^{-|x|^{2}/2},
\end{equation}
where $j, k \in \mN$ and $H_k^{(\ell)} $, $(\ell = 1, \ldots, \dim{\cH_k })$ a basis for $\cH_k$, the space of spherical harmonics of degree $k$. In other words, $\cH_k := \ker \Delta \cap \cP_k$ with $\cP_k$ the space of homogeneous polynomials of degree $k$. The functions $L_j^{\alpha}(t)$ are the Laguerre polynomials. The Hermite functions satisfy
\begin{equation}
\label{FracFTeig}
\cF_{\a}(\phi_{j, k, \ell}) = e^{-i \alpha (2j + k)} \phi_{j, k, \ell},
\end{equation}
which, as a consequence, shows the semigroup property $\cF_{\alpha} \cF_{\beta} =\cF_{\alpha + \beta}$. The reader may wish to keep formulas (\ref{HermFrac}) and (\ref{FracFTeig}) in mind, for comparison with the generalized Fourier transforms we will discuss in the rest of this review.

Obviously, also a formulation {\bf F4} exists for the fractional Fourier transform. This can be obtained by taking the limit $\beta \rightarrow 0$ in the subsequent Theorem \ref{FracSeries}.

Let us end this section by mentioning that there also exist several discrete versions of the fractional Fourier transform. It turns out to be easy to define a fractional version of the discrete FT given by the $n \times n$ matrix $E$. The components of this matrix are determined by $n$-th roots of unity via
\[
E_{j, k} = \frac{1}{\sqrt{n}} e^{-i 2 \pi  \frac{j k}{n}}.
\]
The fractional version of this discrete transform is described in detail in \cite{MR1698479}.  
From the point of view of the present review, the so-called Fourier-Kravchuk transform \cite{MR1456607} is more interesting, as it is also related to a Lie algebra realization  of $\mathfrak{su}_2$ but on a finite dimensional vector space. Moreover, the Fourier-Kravchuk transform is derived from an oscillator model, which is not the case for the fractionalization of $E$.

\subsection{The Dunkl transform}
\label{DTF}

Denote by $\langle .,. \rangle$ the standard Euclidean scalar product in $\mR^{m}$ and by $|x| = \langle x, x\rangle^{1/2}$ the associated norm. For $\alpha \in \mR^{m} - \{ 0\}$, the reflection $r_{\alpha}$ in the hyperplane orthogonal to $\alpha$ is given by
\[
r_{\alpha}(x) = x - 2 \frac{\langle \alpha, x\rangle}{|\alpha|^{2}}\alpha, \quad x \in \mR^{m}.
\]

A root system is a finite subset $R \subset \mR^{m}$ of non-zero vectors such that, for every $\alpha \in R$, the associated reflection $r_{\alpha}$ preserves $R$. We will assume that $R$ is reduced, i.e. $R \cap \mR \alpha = \{ \pm \alpha\}$ for all $\alpha \in R$. Each root system can be written as a disjoint union $R = R_{+} \cup (-R_{+})$, where $R_{+}$ and $-R_{+}$ are separated by a hyperplane through the origin. The subgroup $\cG \subset O(m)$ generated by the reflections $\{r_{\alpha} | \alpha \in R\}$ is called the finite reflection group associated with $R$. We will also assume that $R$ is normalized such that $\langle \alpha, \alpha\rangle = 2$ for all $\alpha \in R$. For more information on finite reflection groups we refer the reader to \cite{Humph}.

A multiplicity function $\kappa$ on the root system $R$ is a $\cG$-invariant function $\kappa: R \rightarrow \mC$, i.e. $\kappa(\alpha) = \kappa(h \alpha)$ for all $h \in \cG$. We will denote $\kappa(\alpha)$ by $\kappa_{\alpha}$.

Fixing a positive subsystem $R_{+}$ of the root system $R$ and a multiplicity function $\kappa$, we introduce the Dunkl operators $T_{i}$ ($i=1, \ldots, m$) associated to $R_{+}$ and $\kappa$ by (see \cite{MR951883, MR1827871})
\[
T_{i} f(x):= \partial_{x_{i}} f(x) + \sum_{\alpha \in R_{+}} \kappa_{\alpha} \alpha_{i} \frac{f(x) - f(r_{\alpha}(x))}{\langle \alpha, x\rangle}, \qquad f \in C^{1}(\mR^{m}).
\]
An important property of the Dunkl operators is that they commute, i.e. $T_{i} T_{j} = T_{j} T_{i}$.

The Dunkl Laplacian is given by $\Delta_{\kappa} = \sum_{i=1}^{m} T_i^2$, or more explicitly by
\[
\Delta_{\kappa} f(x) = \nabla^{2} f(x) + 2 \sum_{\alpha \in R_{+}} \kappa_{\alpha} \left( \frac{\langle \nabla f(x), \alpha \rangle}{\langle \alpha, x \rangle}  - \frac{f(x) - f(r_{\alpha}(x))}{\langle \alpha, x \rangle^{2}} \right)
\]
with $\nabla$ the gradient operator.

If we let $\Delta_{\kappa}$ act on $|x|^2$ we find $\Delta_{\kappa} |x|^2 = 2m + 4 \gamma = 2 \mu$, where $\gamma = \sum_{\alpha \in R_+} \kappa_{\alpha}$. We refer to $\mu$ as the Dunkl dimension, because most special functions related to $\Delta_{\kappa}$ behave as if one would be working with the classical Laplace operator in a space with formal dimension $\mu$. 

We further denote by $\cH_k^{\cD}$ the space of Dunkl-harmonics of degree $k$, i.e. $\cH_k^{\cD} := \ker{\Delta_{\kappa}} \cap \cP_k$. The space of Dunkl-harmonics of degree $k$ has the same dimension as the classical space $\cH_k$ of spherical harmonics of degree $k$.

It is possible to construct an intertwining operator $V_{\kappa}$ connecting the usual partial derivatives $\partial_{x_{j}}$ with the Dunkl operators $T_{j}$ such that $T_{j} V_{\kappa} = V_{\kappa} \partial_{x_{j}}$ (see e.g. \cite{MR1273532}). Note that explicit formulae for $V_{\kappa}$ are only known in a few special cases.

The operators $\Delta_{\k}$, $|x|^{2}$ and $\mE + \frac{\mu}{2}$ satisfy the defining relations of the Lie algebra $\mathfrak{sl}_2$, as was obtained in \cite{He},
\begin{align}
\label{Dunklsl}
\begin{split}
\left[\Delta_{\k}, |x|^2 \right] &= 4 (\mE + \frac{\mu}{2})\\
\left[\Delta_{\k}, \mE + \frac{\mu}{2} \right] &= 2\Delta_{\k}\\
\left[|x|^2, \mE + \frac{\mu}{2}\right] &= -2 |x|^{2}.
\end{split}
\end{align}

The weight function related to the root system $R$ and the multiplicity function $\k$ is given by $w_{\k}(x) = \prod_{\alpha \in R_{+}} |\langle \alpha, x\rangle |^{2 \k_{\alpha}}$.
For suitably chosen functions $f$ and $g$ one then has the following property of integration by parts (see \cite{MR1199124})
\begin{equation}
\int_{\mR^{m}} (T_{i} f) g \;  w_{\k}(x)dx = -\int_{\mR^{m}} f \left(T_{i}g\right) w_{\k}(x)dx
\label{skewnessDunkl}
\end{equation}
with $dx$ the Lebesgue measure.

Following the formulation {\bf F2} of the ordinary Fourier transform, we can now introduce a generalization related to the set of Dunkl operators $T_{i}$ (see \cite{deJ}). This so-called Dunkl transform $\mathcal \cF_{k}: L^1(\mR^m, w_{\k}(x)dx)\to C(\mR^m)$ is defined as follows
\[
 \mathcal \cF_{\k} (f)(y):= c_{\k}^{-1} \int_{\mR^m} D(x,y)\,f(x)\, w_{\k}(x) dx
\]
with $c_{\k} = \int_{\mR^m} e^{-|x|^{2}/2}w_{\k}(x) dx$ the Mehta constant related to $\cG$ and where $D(x,y)$ is the Dunkl kernel. This kernel is the unique solution of the system
\[
T_{j, x} D(x,y) = -i y_{j} D(x,y), \quad j=1, \ldots, m
\]
which is real-analytic in $\mR^{m}$ and satisfies $D(0,y)=1$. One can prove that $D(x,y)$ is bounded (see \cite{deJ}), so the Dunkl transform is well defined.

The eigenfunctions of this transform are studied in e.g. \cite{MR1199124, MR1620515}. They are given by
\begin{equation}
\label{HermDunkl}
\phi_{j, k, \ell}^{\k} := L_{j}^{\frac{\mu}{2} + k-1}(|x|^2) H_k^{(\ell)} \, e^{-|x|^{2}/2},
\end{equation}
with $j, k \in \mN$ and $H_k^{(\ell)} $, $(\ell = 1, \ldots, \dim{\cH_k^{\cD} })$ a basis for $\cH_k^{\cD} $. They satisfy $\cF_{\k}(\phi_{j, k, \ell}^{\k}) = (-i)^{2j + k} \phi_{j, k, \ell}^{\k}$. Note that for $\k =0$, these functions reduce to formula (\ref{HermFrac}).

The exponential expression {\bf F3} takes for the Dunkl transform the following form
\[
\cF_{\k} = e^{ \frac{i \pi \mu}{4}} e^{\frac{i \pi}{4}(\Delta_{\k} - |x|^{2})},
\]
as was shown in \cite{Said}. The operator $H_{\k} =-(\Delta_{\k} - |x|^{2})/2$ describes a quantum system of Calogero-Moser-Sutherland type. We refer the reader to \cite{vD} for a detailed description of such systems and to \cite{MR1620515} for a solution of the spectral problem using Dunkl operators. Also an expansion of the type {\bf F4} has been obtained, see Proposition 2.4 in \cite{MR1973996} .

In the one-dimensional case, there exists only one reflection group, namely $\mZ_{2}$. In that case, the kernel of the Dunkl transform is explicitly known and the transform is given by
\[
\cF_{\k}(f)(y) = \frac{1}{ 2^{\frac{2\k+1}{2}}} \int_{-\infty}^{ \infty} \left(  \widetilde{J}_{\k-\frac{1}{2}}\left(  |x y|\right) -i\frac{x y}{ 2}\widetilde{J}_{\k+\frac{1}{2}}\left(  |x y| \right)  \right) f(x) |x|^{2\k} dx
\]
where $\k \geq -1/2$ is the multiplicity parameter and with $\widetilde{J}_{\nu}(z) = (z/2)^{-\nu}J_{\nu}(z)$. For a detailed treatment, in particular its relation to the Hankel transform, we refer the reader to \cite{C}. For $\k=0$, this transform reduces to the ordinary Fourier transform.

It is interesting to note that this one dimensional transform reappears in many different contexts. In the context of Wigner quantization, this kernel appeared first in e.g. \cite{MR0585590, MR0604452}. In related work, a further classification of all $\mathfrak{osp}(1|2)$ representation spaces and the associated quantum systems has been obtained in \cite{MR2779617}. Similar to the classical Fourier transform, there also exists a finite oscillator model that in the continuum limit yields the Dunkl transform. This is the so-called Fourier-Hahn transform introduced recently in \cite{J}.

For higher values of the dimension, no explicit formula for the Dunkl kernel is known, except for a few special cases. Note that in the higher dimensional case, it is easy to prove that there no longer exists a connection with Wigner quantization (for the Lie superalgebra $\mathfrak{osp}(1|2n)$, see e.g. \cite{J1}), not even when considering the reflection group obtained by taking the product of several copies of $\mZ_{2}$.

\subsection{The radially deformed Fourier transform}
\label{radFT}

It turns out that the $\mathfrak{sl}_{2}$ relations (\ref{Dunklsl}) also hold for the generalized operators $|x|^a$, $|x|^{2-a} \Delta_{\k}$ and $\mE + \frac{a+\mu-2}{2}$, with $a >0$ a real parameter. Indeed, one has
\begin{align*}
\left[|x|^{2-a}\Delta_{\k}, |x|^a \right] &= 2 a \, (\mE + \frac{a +\mu -2}{2})\\
\left[|x|^{2-a}\Delta_{\k}, \mE + \frac{a+\mu-2}{2} \right] &= a\, |x|^{2-a}\Delta_{\k}\\
\left[|x|^a, \mE + \frac{a +\mu-2}{2}\right] &= -a \,|x|^a.
\end{align*}
This was first observed, in the context of minimal representations, for $a=1$ and $\k =0$ in \cite{MR2134314, MR2401813} and subsequently generalized to arbitrary $a$ and $\k$ in \cite{Orsted2}. 

The paper \cite{Orsted2} was mostly concerned with the study of the associated Hermite semigroup given by
\[
\cJ_{\k,a}(\o):=e^{\frac{\o}{a} \left( |x|^{2-a}\Delta_{\k} - |x|^a\right)}
\]
where $\o$ is a complex parameter satisfying $\Re{\o} \geq 0$. This semigroup was studied in great detail and in particular an integral operator expression was found where the kernel is given as a series expansion. An important tool was the construction of an eigenbasis for the hamiltonian $H_{\k, a} =- \left( |x|^{2-a}\Delta_{\k} - |x|^a\right)/a$. Putting, for $j, k \in \mN$ and $H_k^{(\ell)} $, $(\ell = 1, \ldots, \dim{\cH_k^{\cD} })$ a basis for $\cH_k^{\cD} $,
\begin{equation}
\label{HermRad}
\phi_{j, k, \ell}^{\k,a} := L_{j}^{\frac{\mu + 2k-2}{a}}\left(\frac{2}{a}|x|^a\right) H_k^{(\ell)} \, e^{-|x|^a/a},
\end{equation}
lengthy computations show
\[
H_{\k,a} \phi_{j, k, \ell}^{\k,a} = \left( \frac{\mu - 2}{a} + \frac{2k}{a}+2j+1\right) \phi_{j, k, \ell}^{\k,a}.
\]
The set of functions $\{ \phi_{j, k, \ell}^{\k,a} \}$ forms an orthogonal basis for the space $L^2(\mR^m, \vartheta_{\k,a}(x)dx)$ with $\vartheta_{\k,a}(x) = |x|^{a-2}w_{\k}(x)$.

In order to keep the subsequent formulas as simple as possible, we restrict ourselves in this review to the case where $\k=0$ and to the specific semigroup parameter $\o= i \pi /2$. This yields the so-called radially deformed Fourier transform
\[
\cF_{0,a} = e^{ \frac{i \pi (m+a-2)}{2a}} e^{\frac{i \pi}{2 a}(|x|^{2-a}\Delta - |x|^{a})},
\]
where a suitable normalization has been added to make the transform unitary. A series expansion of its integral kernel is given in the subsequent theorem, which was obtained in \cite{Orsted2}.

\begin{theorem}
Put
\[
K_{a}(x,y) = a^{2 \lambda/a} \Gamma \left( \frac{2 \lambda+a}{a}\right)\sum_{k=0}^{\infty} e^{-\frac{i \pi k}{a}}  \frac{\lambda +k}{\lambda} z^{-\lambda} J_{\frac{2(k+ \lambda)}{a}}\left( \frac{2}{a} z^{a/2}\right) \; C_{k}^{\lambda}(w),
\]
with $\lambda =(m-2)/2$, $z = |x| |y|$ and $w = \la x, y \ra /z$. This series is convergent and the integral transform 
\[
\cF_{0,a} (f)(y) = \frac{\Gamma(m/2)}{ \Gamma (\frac{2 \lambda+a}{a}) 2 a^{2\lambda/a} \pi^{m/2}} \int_{\mR^{m}} K_{a}(x,y) f(x) \vartheta_{0,a}(x)dx
\]
defined on the function space  $L^2(\mR^m, |x|^{a-2}dx)$  coincides with the operator $\cF_{0,a} = e^{ \frac{i \pi (m+a-2)}{2a}} e^{\frac{i \pi}{2 a}(|x|^{2-a}\Delta - |x|^{a})}$ on the basis $\{ \phi_{j, k, \ell}^{0,a} \}$:
\[
\cF_{0,a}  \left(\phi_{j, k, \ell}^{0,a}  \right)= e^{-i \pi (j + \frac{k}{a})}\phi_{j, k, \ell}^{0,a}.
\]
\end{theorem}

This theorem can be proven using the integral identity (see \cite[exercise 21, p. 371]{Sz})
\[
\int_{0}^{+\infty} r^{\alpha+1}  J_{\a}(rs)\,  L_{j}^{\a}(r^{2}) e^{-r^{2}/2}dr = (-1)^{j}s^{\a} L_{j}^{\a}(s^{2}) e^{-s^{2}/2}
\]
combined with the fact that the Gegenbauer polynomials yield the reproducing kernel for spaces of spherical harmonics (see formula (\ref{FH}) in Section \ref{prelims}).

Note that for $a=2$, the kernel $K_{a}(x,y)$ reduces to the usual exponential kernel of the ordinary Fourier transform, see {\bf F4}. Also when $a=1$, a closed form is known, given by
\[
K_{1}(x,y) = \Gamma \left(\frac{m-1}{2}\right)  \widetilde{J}_{\frac{m-3}{2}} \left( \sqrt{2 (|x| |y| + \la x, y \ra )} \right).
\]
This result was already obtained in \cite{MR2401813} using a geometric construction.

For arbitrary $a$, such a closed form is not available. Moreover, there are no bounds known on $K_{a}(x,y)$ for $a \neq 1$ or $2$. Also a formulation {\bf F2}, characterizing the kernel $K_{a}(x,y)$ as the unique eigenfunction of a system of PDEs is not known. Note that in recent work \cite{HDBradial}, substantial progress has been made in determining the kernel for $a=2/n$, $n \in \mN$.

To conclude this section, we consider the radially deformed Fourier transform in dimension 1, for the reflection group $\mZ_2$ (i.e. the general rank one case). Then the kernel is again known explicitly for arbitrary $a$ and $\k$ (where $\k$ is now a real number with $2 \k > 1-a$) and given by
\[
\cF_{\k,a}(f)(y) = \frac{1}{2 a^{\frac{2\k-1}{a}}} \int_{-\infty}^{ \infty} \left(  \widetilde{J}_{\frac{2\k-1}{a}}\left( \frac{2}{a} |x y|^{\frac{a}{2}}\right) +\frac{x y}{ (a i)^{\frac{2}{a}}} \widetilde{J}_{\frac{2\k+1}{a}}\left( \frac{2}{a} |x y|^{\frac{a}{2}}\right)  \right) f(x) |x|^{2\k+a-2} dx.
\]
Up to now, no discrete counterpart of this transform has been obtained.

\subsection{Fourier transform in superspace}

In this section we give a brief overview of results obtained concerning the Fourier transform in superspace. Again, this generalization was obtained via an extension of the $\mathfrak{sl}_2$ relations in formula (\ref{sl2relclass}).

We start by considering the flat supermanifold $\mR^{m|2n}=(\mR^m,\cC^\infty_{\mR^m}\otimes \Lambda_{2n})$ with $m$ bosonic variables $x_i$ and $2n$ fermionic variables ${x\grave{}}_j$ (generating the Grassmann algebra $\Lambda_{2n}$), equipped with an orthosymplectic metric. The even number of fermionic variables is needed to allow the symplectic metric. The supervector ${\bf x}$ is defined as
\[
{\bf x}=(X_1,\cdots,X_{m+2n})=(x_1,\cdots,x_m,{x\grave{}}_1,\cdots,{x\grave{}}_{2n}).
\]
Unless stated otherwise we will always assume $m\not=0$.

The inner product of two supervectors ${\bf x}$ and ${ \bf y}$ is given by
\begin{equation}
\label{inprod}
\langle {\bf x},{\bf y}\rangle=\sum_{i=1}^mx_iy_i-\frac{1}{2}\sum_{j=1}^n({x\grave{}}_{2j-1}{y\grave{}}_{2j}-{x\grave{}}_{2j}{y\grave{}}_{2j-1}).
\end{equation}
The commutation relations for two supervectors are determined by $X_iY_j=(-1)^{[i][j]}Y_jX_i$ with $[i]$ given by
\[
[i]=0\quad \mbox{if $i\le m$} \qquad \mbox{and} \qquad [i]=1\quad \mbox{otherwise}.
\]
This implies that the ${x\grave{}}_{i}$ and ${y\grave{}}_{i}$ together generate the Grassmann algebra $\Lambda_{4n}$. This also means that the inner product \eqref{inprod} is symmetric, i.e. $\langle {\bf x},{\bf y}\rangle=\langle{\bf y},{\bf x}\rangle$.  Note that the inner product can equivalently be written as $\langle {\bf x},{\bf y}\rangle = \sum_{ij}X_ig^{ij}Y_j$ with metric $g$ defined as
\[
\begin{cases} 
g^{ii}=1 & 1\le i\le m,\\
g^{2i-1+m,2i+m}=-1/2 & 1\le i\le n ,\\  
g^{2i+m,2i-1+m}=1/2 & 1\le i\le n,\\
g^{ij}=0 & \mbox{otherwise}. 
\end{cases}
\]

We put $X^j=\sum_iX_ig^{ij}$, so for the commuting variables $x^j=x_j$ and for the anticommuting variables ${x\grave{}}^{2j-1}=\frac{1}{2}{x\grave{}}_{2j}$ and ${x\grave{}}^{2j}=-\frac{1}{2}{x\grave{}}_{2j-1}$ holds. Considering the symmetry of the metric, $g^{ij}=(-1)^{[i]}g^{ji}$, we subsequently obtain 
\[
\langle {\bf x},{\bf y}\rangle=\sum_jX^jY_j=\sum_{i,j}X_ig^{ji}(-1)^{[i]}Y_j=\sum_i(-1)^{[i]}X_iY^i.
\]
The generalized norm squared is given by
\begin{equation*}
R^2=\langle {\bf x},{\bf x}\rangle=\sum_{j=1}^{m+2n}X^jX_j =\sum_{i=1}^mx_i^2-\sum_{j=1}^n{x\grave{}}_{2j-1}{x\grave{}}_{2j}.
\end{equation*}

The fermionic partial derivatives $\partial_{{x\grave{}}_{j}}$ commute with the bosonic variables and satisfy the Leibniz rule $\partial_{{x\grave{}}_{j}}{x\grave{}}_{k}=\delta_{jk}-{x\grave{}}_{k}\partial_{{x\grave{}}_{j}}$. The super gradient is defined as $\nabla=(\partial_{X^1},\cdots,\partial_{X^{m+2n}})$. Using $\partial_{{x\grave{}}^{2j-1}}=2\partial_{{x\grave{}}_{2j}}$ we find $\nabla=(\partial_{x_1},\cdots,\partial_{x_m},2\partial_{{x\grave{}}_{2}},-2\partial_{{x\grave{}}_{1}},\cdots,2\partial_{{x\grave{}}_{2n}},-2\partial_{{x\grave{}}_{2n-1}})$ and $\nabla^j=(-1)^{[j]}\partial_{X_j}$. The super Laplace operator is given by
\begin{align*}
\nabla^2=\langle \nabla,\nabla \rangle&=\sum_{k=1}^{m+2n}\nabla^k\nabla_k=\sum_{i=1}^m\partial_{x_i}^2-4\sum_{j=1}^n\partial_{{x\grave{}}_{2j-1}}\partial_{{x\grave{}}_{2j}}.
\end{align*}
The super Euler operator is defined as
\[
\mE=\langle {\bf x},\nabla\rangle=\sum_{k=1}^{m+2n}X^k\nabla_{k}=\sum_{k=1}^{m+2n}X_k\partial_{X_k}.
\]

We can now state the important theorem yielding a new realization of the Lie algebra  $\mathfrak{sl}_2$. This result was obtained first in \cite{DBS5}.
\begin{theorem}
\label{sl2lemma}
The operators $\nabla^2/2$, $R^2/2$ and $\mE+M/2$ on $\mR^{m|2n}$ for each two integers $m,n$, with $M=m-2n$, generate $\mathfrak{sl}_2$. This is a consequence of the commutators 
\begin{align*}
\left[\nabla^2/2,R^2/2\right]&=\mE+M/2\\
\left[\nabla^2/2,\mE+M/2\right]&=2\nabla^2/2\\
\left[R^2/2,\mE+M/2\right]&=-2R^2/2.
\end{align*}
\end{theorem}

The parameter $M=m-2n$ will often give a global characterization of the superspace $\mR^{m|2n}$ because of its appearance in the $\mathfrak{sl}_2$-relations. For that reason $M$ is sometimes called the superdimension. Also note that the three operators in the theorem are invariant under the Lie superalgebra $\mathfrak{osp}(m|2n)$, see e.g. \cite{CLie}.

The space of superpolynomials is given by $\cP=\mR[x_1,\cdots,x_m]\otimes \Lambda_{2n}$. In general, for a function space $\cT$ corresponding to the $m$ bosonic variables (e.g. $\cS(\mR^{m})$, $L^p(\mR^m)$, $C^k(\Omega)$) one uses the notation $\cT_{m|2n}=\cT\otimes\Lambda_{2n}$. The null-solutions of the super Laplace operator are called harmonic superfunctions. In particular we are interested in harmonic superpolynomials. 
An element $F \in \cP$ is a spherical harmonic of degree $k$ if it satisfies
\[
\nabla^2 F =0 \qquad  \mbox{and} \qquad \mE F = kF. 
\]
The space of all spherical harmonics of degree $k$ is denoted by $\cH_k^{m | 2n}$. For a detailed study of super harmonics, we refer the reader to \cite{DBS5, DBE1}.

The integration used on $\Lambda_{2n}$ is the so-called Berezin integral (see \cite{MR732126}), defined by
\[
\int_{B} := \pi^{-n} \partial_{{x \grave{}}_{2n}} \ldots \partial_{{x \grave{}}_{1}}.
\]
On a general flat superspace $\mR^{m|2n}$  the integration is subsequently defined by
\[
\int_{\mR^{m | 2n}} := \int_{\mR^m} d x\int_B=\int_B \int_{\mR^m} d x,
\]
with $d x$ the usual Lebesgue measure in $\mR^{m}$. Note that in the super case we omit the measure. This expression is defined on the product of the space of Lebesgue integrable functions with the Grassmann algebra.

The super Fourier transform on $\cS(\mR^m)_{m|2n}$ was first introduced in \cite{DBS9} as
\[
\cF^{\pm}_{m|2n}(f)({\bf y})=(2\pi)^{-\frac{M}{2}}\int_{\mR^{m|2n},{\bf x}}\exp(\pm i\langle {\bf x},{\bf y}\rangle)f({\bf x}),
\]
with $\langle {\bf x},{\bf y}\rangle$ given in \eqref{inprod}, yielding a $\mathfrak{osp}(m|2n)$-invariant (and hence also an $O(m)\times Sp(2n)$-invariant) generalization of the purely bosonic Fourier transform. This transform clearly satisfies  $\cF^{\pm}_{m|2n}=\cF^{\pm}_{m|0}\circ\cF^{\pm}_{0|2n}$.

Again the kernel $K_{\pm}( {\bf x},{\bf y}) = \exp(\pm i\langle {\bf x},{\bf y}\rangle)$ is given as the unique solution, up to a multiplicative constant, of a system of PDEs:
\[
\partial_{X^j} K_{\pm}( {\bf x},{\bf y}) = \pm i \,Y_j \,K_{\pm}( {\bf x},{\bf y}), \qquad j = 1, \ldots, m+ 2n
\]
with initial condition $K_{\pm}( 0,{\bf y})=1$, thus generalizing the characterization {\bf F2} of the ordinary Fourier transform.

 An important property of the super Fourier transform is
\[
\cF_{m|2n}^\pm\left(\nabla^2f\right)({\bf y})=-R^2_{{\bf y}}\cF_{m|2n}^\pm \left(f\right)({\bf y}).
\]
In this formula we introduced the notation $R^2_{{\bf y}}=\langle \by,\by\rangle$ for super vector variables other than $\bx$.

The super (spherical) Hermite functions were introduced in \cite{MR2371128} as
\begin{equation}
\label{HermSuper}
\varphi_{j,k,\ell} := 2^{2j} j! L^{\frac{M}{2}+k-1}_{j}(R^2)   H_k^{(\ell)} \exp{(-R^2/2)}
\end{equation}
where $j, k \in \mN$
and with $H_k^{(\ell)} $, $(\ell = 1, \ldots, \dim{\cH_k^{m | 2n}})$ a basis for $\cH_k^{m | 2n}$. In case $M\not\in-2\mN$ or $m=0$ they constitute a basis for the vector space $\cP\otimes\exp(-R^2/2)$ and via density also for  $\cS(\mR^m)_{m|2n}$. Note that orthogonality relations also exist for these functions, see \cite{CPLMS}, although they are difficult to obtain.

After tedious computations, it was obtained in \cite{DBS9} that these generalized Hermite functions are eigenfunctions of the Fourier transform, i.e.
\begin{equation}
\label{FourCH}
\cF^{\pm}_{m|2n}(\varphi_{j,k,\ell})({\bf y})=  (\pm i) ^{2j+k}\varphi_{j,k,\ell}({\bf y}).
\end{equation}
Moreover, as observed in \cite{DBS9} and formally proven in \cite{CDB}, equation (\ref{FourCH}) leads to an exponential form for the super Fourier transform, given by
\[
\cF^\pm_{m|2n}=\exp\left(\mp\frac{i\pi M}{4}\right)\,\exp\left(\pm\frac{i\pi}{4}(R^2-\nabla^2)\right).
\]
Note that $H = -(\nabla^2-R^2)/2$ is the hamiltonian of the super harmonic oscillator, describing a system of $m$ bosonic and $2n$ fermionic particles.

Also formulation {\bf F4} has been obtained in this case. The following expansion holds
\[
\exp(\pm i\langle {\bf x},{\bf y}\rangle)= 2^{M/2-1}\Gamma(M/2)  \sum_{k=0}^\infty (\pm i)^k \frac{2k+M-2}{M-2}  \frac{J_{\frac{M}{2}+k-1}(R_{{\bf x}}R_{{\bf y}})}{(R_{{\bf x}}R_{{\bf y}})^{\frac{M}{2}+k-1}} (R^2R_{{\bf y}}^2)^{(k/2)} C^{(M-2)/2}_k \left(\frac{\langle {\bf x},{\bf y} \rangle}{(R^2R_{{\bf y}}^2)^{(1/2)}} \right).
\]
The proof of this statement is quite involved and was given in \cite{MR2683546}, based on results in \cite{CPLMS}. In particular, one needs to take care in the proper definition of a function depending on $R_x$ or $R_y$, see \cite{MR2683546}.

There also exists a fractional version of the super Fourier transform. Indeed, when $M\not\in-2\mN$ or when $m=0$, we define the operator $\cF^{\alpha}_{m|2n}$ on $\cS(\mR^m)\otimes\Lambda_{2n}$,  by its action on the basis functions: 
\[
\cF^{\alpha}_{m|2n} (\varphi_{j,k,\ell})({\bf y}) = e^{  i \alpha (2j+k)} \varphi_{j,k,\ell}({\bf y}),
\]
where $\alpha \in [-\pi,\pi]$. The fractional Fourier transform thus rotates the basis functions over a multiple of the angle $\alpha$. In the limit case $\alpha = \pm \pi/2$, the fractional Fourier transform reduces to the super Fourier transform, i.e. $\cF^{\pm \pi/2}_{m|2n} = \cF^{\pm}_{m|2n}$. We have the following integral representation, see \cite{DBS9}.

\begin{theorem}
On $\cS(\mR^m)_{m|2n}$, the fractional Fourier transform is given by the following integral operator
\[
\cF^{\alpha}_{m|2n}(f)({\bf y})= \left(\pi (1- e^{2i \alpha})\right)^{-M/2} \int_{\mR^{m|2n},x}   \exp{\left(\frac{ 4 e^{i \alpha} \langle {\bf x},{\bf y} \rangle - (1+ e^{2i \alpha})(R^2 + R_{{\bf y}}^2)}{2- 2e^{2i \alpha}}\right)} f({\bf x})
\]
when $\alpha \in ]-\pi,\pi[$ with $\alpha \neq 0$.
\end{theorem}

Let us conclude this section by mentioning a few applications of the super Fourier transform. It has been used to give an alternative proof for the fundamental solution of the super Laplace operator, obtained in \cite{MR2386499}. It has also been used to solve the super Schr\"odinger equation with delta potential \cite{DBS8}. Finally, it plays an important role in the study of the Hilbert space for quantum mechanics in superspace \cite{CDB}.

\begin{remark}
It is also possible to define generalized Fourier transforms in the context of supermanifolds using a completely different strategy. This is e.g. done for Heisenberg-Clifford Lie supergroups in \cite{All}.
\end{remark}

\begin{remark}
There also exist versions of the Fourier transform related to other generalized geometries, that are defined using similar ideas. In particular, we mention the $q$ deformed Fourier transform, which is related to $\cU_q(\mathfrak{sl}_2)$. This is studied in e.g. \cite{Koorn} in the one dimensional case and in \cite{CSigma} for general dimension.
\end{remark}

\section{Hypercomplex Fourier transforms}

In this section, we generalize the framework for scalar Fourier transforms related to $\mathfrak{sl}_2$, as discussed in Section \ref{Secsl}, to hypercomplex Fourier transforms. These are related to the Lie superalgebra $\mathfrak{osp}(1|2)$, which contains $\mathfrak{sl}_2$ as its even subalgebra. In order to do so, we need to introduce the framework of Clifford analysis (Section \ref{prelims}), as we now need to work with (generalized) Dirac operators instead of Laplace operators.

Again we will discuss various examples of generalized Fourier transforms in this context. We will focus our attention in particular on the Clifford-Fourier transform (Section \ref{CFTsection}) and the radially deformed hypercomplex Fourier transform (Section \ref{HRFTsection}). Finally, in Section \ref{OtherTFsSect} we summarize some results on other hypercomplex Fourier transforms that do not fit into the $\mathfrak{osp}(1|2)$ framework.

\subsection{Preliminaries on Clifford analysis}
\label{prelims}

The Clifford algebra $\cC l_{0, m}$ over $\mR^{m}$ is the algebra generated by $e_{i}$, $i= 1, \ldots, m$, under the relations
\begin{align} \label{eq:eij}
\begin{split}
&e_{i} e_{j} + e_{j} e_{i} = 0, \qquad i \neq j,\\
& e_{i}^{2} = -1.
\end{split}
\end{align}
This algebra has dimension $2^{m}$ as a vector space over $\mR$. It can be decomposed as $\cC l_{0, m} = \oplus_{k=0}^{m} \cC l_{0, m}^{k}$
with $\cC l_{0, m}^{k}$ the space of $k$-vectors defined by
\[
\cC l_{0, m}^{k} := \mbox{span} \{ e_{i_{1}} \ldots e_{i_{k}}, i_{1} < \ldots < i_{k} \}.
\]
The projection on the space of $k$-vectors is denoted by $[\, .\, ]_{k}$.

The operator $\bar{.}$ is the main anti-involution on the Clifford algebra $\cC l_{0, m}$ defined by
\[
\overline{a b} = \overline{b} \overline{a}, \qquad \overline{e_{i}} = -e_{i}, \quad (i = 1,\ldots, m).
\]
Similarly we have the automorphism $\epsilon$ given by
\[
\epsilon(a b) = \epsilon(a)\epsilon( b), \qquad \epsilon(e_{i}) = -e_{i}, \quad (i = 1,\ldots, m).
\]

In the sequel, we always consider functions $f$ taking values in $\cC l_{0, m}$, unless explicitly mentioned. Such functions can be decomposed as
\begin{equation} \label{clifford_func}
f(x) = f_{0}(x) + \sum_{i=1}^{m} e_{i}f_{i}(x) + \sum_{i< j} e_{i} e_{j} f_{ij}(x) + \ldots + e_{1} \ldots e_{m} f_{1 \ldots m} (x)
\end{equation}
with $f_{0}, f_{i}, f_{ij}, \ldots, f_{1 \ldots m}$ all real- or complex-valued functions on $\mR^m$.

Several important groups can be embedded in the Clifford algebra. Note that the space of $1$-vectors in $\cC l_{0, m}$ is canonically isomorphic to $\mR^{m}$. Hence we can define
\[
Pin(m) = \left\{ s_{1} s_{2} \ldots s_{n} | n \in \mN, s_{i} \in \cC l_{0, m}^{1} \mbox{ such that }  s_{i}^{2}=-1 \right\},
\]
i.e., the Pin group is the group of products of unit vectors in $\cC l_{0, m}$. This group is a double cover of the orthogonal group $O(m)$ with covering map $p: Pin(m) \rightarrow O(m)$, which we will describe explicitly in Section \ref{HRFTsection}.

Similarly we define
\[
Spin(m) = \left\{ s_{1} s_{2} \ldots s_{2n} | n \in \mN, s_{i} \in \cC l_{0, m}^{1} \mbox{ such that }  s_{i}^{2}=-1 \right\},
\]
i.e., the Spin group is the group of even products of unit vectors in $\cC l_{0, m}$.
This group is a double cover of $SO(m)$. For more information about Clifford algebras and analysis, we refer the reader to \cite{MR697564, MR1169463, GM}.

The Dirac operator is given by $\upx := \sum_{j=1}^{m} \partial_{x_{j}} e_{j}$ and the vector variable by $\ux := \sum_{j=1}^{m} x_{j} e_{j}$.
The square of the Dirac operator equals, up to a minus sign, the Laplace operator in $\mR^{m}$: $\upx^{2} = - \Delta$. Together, the Dirac operator and the vector variable generate the Lie superalgebra $\mathfrak{osp}(1|2)$. This is the subject of the following theorem. It is not entirely clear where this result was first obtained. One of the earliest instances seems to be \cite{HS}.

\begin{theorem}
The operators $\upx$ and $\ux$ generate a Lie superalgebra, isomorphic with $\mathfrak{osp}(1|2)$, with the following relations
\begin{equation}
\begin{array}{lll}
\{ \ux, \ux \} = -2 |x|^{2} & \quad & \{ \upx, \upx \} = -2 \Delta\\
\vspace{-3mm}\\
\{ \ux, \upx\} = -2 \left( \mE + \frac{m}{2}\right)&\quad&\left[\mE + \frac{m}{2}, \upx \right] = - \upx\\
\vspace{-3mm}\\
\left[ |x|^{2}, \upx \right] = -2\ux&\quad&\left[\mE + \frac{m}{2}, \ux \right] =   \ux\\
\vspace{-3mm}\\
\left[ \Delta, \ux \right] = 2\upx&\quad&\left[\mE + \frac{m}{2}, \Delta \right] = - 2 \Delta\\
\vspace{-3mm}\\
\left[ \Delta^{2}, |x|^{2} \right] = 4 \left( \mE + \frac{m}{2} \right)&\quad&\left[\mE + \frac{m}{2}, |x|^{2} \right] = 2 |x|^{2},
\end{array}
\end{equation}
where $\mE = \sum_{i=1}^{m}x_{i}\partial_{x_{i}}$ is the Euler operator.
\label{ospFamily}
\end{theorem}

The classical Laplace operator $\Delta$ and $|x|^{2}$ together generate $\mathfrak{sl}_{2}$, see formula (\ref{sl2relclass}), and this Lie algebra forms the even subalgebra of $\mathfrak{osp}(1|2)$. 

We further introduce the so-called Gamma operator (see e.g. \cite{MR1169463})
\[
\Gamma_{x} := - \sum_{j<k} e_{j}e_{k} (x_{j} \partial_{x_{k}} - x_{k}\partial_{x_{j}}) = -\ux \upx - \mE. 
\]
Note that $\Gamma_{x}$ commutes with radial functions, i.e. $[\Gamma_{x}, f(|x|)] =0$.

Denote by $\cP\otimes \cC l_{0, m}$ the space of polynomials taking values in $\cC l_{0, m}$, i.e. 
$$
   \cP\otimes \cC l_{0, m} : = \mR[x_{1}, \ldots, x_{m}] \otimes \cC l_{0, m}.
$$ 
The space of homogeneous, Clifford-valued polynomials of degree $k$ is then denoted by $\cP_{k} \otimes \cC l_{0, m}$. 
The space $\cM_{k} : = \ker{\upx} \cap \left( \cP_{k} \otimes \cC l_{0, m}\right)$ is called the space of spherical monogenics 
of degree $k$. Similarly,  $\cH_{k} \otimes \cC l_{0, m}:= \ker{\Delta} \cap \left( \cP_{k}\otimes \cC l_{0, m}\right)$ is the space of (Clifford-valued) spherical 
harmonics of degree $k$. 

The elements of $\cH_k\otimes \cC l_{0, m}$ are functions of the form \eqref{clifford_func}
with $f_{i_1,\ldots,i_j}$ being ordinary harmonics. It follows, in particular, that the reproducing kernel
of $\cH_k\otimes \cC l_{0, m}$ is $\frac{\lambda+k}{\l} C^{\lambda}_{k}(\langle \xi,\eta \rangle)$, with $\l= (m-2)/2$ and 
$C_k^\lambda$ being the Gegenbauer polynomial, the same reproducing kernel for the space of 
ordinary spherical harmonics of degree $k$ (cf. \cite{MR1827871,V}). This means that
\begin{equation} \label{FH}
\frac{\lambda+k}{\l} \int_{\mS^{m-1}} C^{\lambda}_{k}(\langle \xi,\eta \rangle) H_{\ell}(\xi) d \sigma(\xi) = \sigma_m \, \delta_{k \ell} \, H_{\ell}(\eta), \quad H_{\ell} \in \cH_{\ell}\otimes \cC l_{0, m}
\end{equation}
with $\sigma_m = 2 \pi^{m/2}/\Gamma(m/2)$. The definition shows immediately that $\cM_{k} \subset \cH_{k}\otimes \cC l_{0, m}$. More precisely, we have the following Fischer decomposition (see \cite[Theorem 1.10.1]{MR1169463}):
\begin{equation}
\cH_{k}\otimes \cC l_{0, m} = \cM_{k} \oplus \ux \cM_{k-1}.
\label{Fischer}
\end{equation}
It is easy to construct projection operators to the components of this decomposition. They are given by 
(\cite[Corollary 1.3.3]{MR1169463})
\begin{align}
\label{projGamma}
\begin{split}
\mP_{1} &= 1+\frac{\ux \upx}{2k+m-2}, \\
\mP_{2} &= -\frac{\ux \upx}{2k+m-2},
\end{split}
\end{align}
and satisfy $\mP_{1}+ \mP_{2} = 1$, $\mP_{1} (\cH_{k}\otimes \cC l_{0, m} )= \cM_{k}$ and $\mP_{2}(\cH_{k} \otimes \cC l_{0, m})= \ux \cM_{k-1}$.
We also have the relations 
\begin{align} \label{gammaeig1}
\begin{split}
\Gamma_{x} \cM_{k} &= -k \cM_{k},\\
\Gamma_{x} ( \ux \cM_{k-1})&= (k+m -2) \ux \cM_{k-1},
\end{split}
\end{align}
which follows easily from $\Gamma_{x} = -\ux \upx - \mE$ and Theorem \ref{ospFamily}.

Also the space of monogenic polynomials has a reproducing kernel, which can e.g. be obtained using the Funk-Hecke theorem, see \cite{DBOSS}.

\begin{proposition}
\label{reprkernels}
Put $x = r x'$ and $y =s y'$ with $x', y' \in \mS^{m-1}$. Furthermore, put $\lambda = (m-2)/2$ and $\sigma_m = 2 \pi^{m/2}/\Gamma(m/2)$. 
For $k\in \mN^{*}$ put
\begin{align*}
P_{k}(x', y') &= \frac{k+2\l}{2\l} C_{k}^{\l}(\la x' , y' \ra) -(\ux' \wedge \uy' ) C_{k-1}^{\lambda+1}(\la x' , y' \ra),\\ 
Q_{k-1}(x', y') &= \frac{k}{2\l} C_{k}^{\l}(\la x' , y' \ra) +(\ux' \wedge \uy' ) C_{k-1}^{\lambda+1}(\la x' , y' \ra)
\end{align*}
with $P_{0}(x', y')=C_{0}^{\l}(0)=1$. Then, for $M_{\ell} \in \cM_{\ell}$, one has
\[
\begin{array}{ll} \int_{\mS^{m-1}} P_{k}(x', y') M_{\ell}(x')d \s (x') = \sigma_m \delta_{k, \ell} M_{\ell}(y')\\
\\
\int_{\mS^{m-1}} P_{k}(x', y') \ux'M_{\ell}(x')d \s (x') = 0
\end{array} 
\quad \mbox{and} \quad
\begin{array}{ll} \int_{\mS^{m-1}} Q_{k-1}(x', y') M_{\ell}(x')d \s (x') = 0\\
\\
\int_{\mS^{m-1}} Q_{k-1}(x', y') \ux'M_{\ell}(x')d \s (x') =  \sigma_m \delta_{k, \ell+1} y'M_{\ell}(y').
\end{array} 
\]
\end{proposition}

Next we define the inner product and the wedge product of two vectors $\ux$ and $\uy$
\begin{align*}
\langle x, y  \rangle &:= \sum_{j=1}^{m} x_{j} y_{j} = - \frac{1}{2} (\ux \,\uy + \uy \, \ux)\\
\ux \wedge \uy &:= \sum_{j<k} e_{j}e_{k} (x_{j} y_{k} - x_{k}y_{j}) = \frac{1}{2} (\ux \,\uy - \uy \, \ux).
\end{align*}
For the sequel we need the square of $\ux \wedge \uy$. A short computation (see \cite{DBXu}) shows
\begin{align*}
(\ux \wedge \uy )^{2} &=  - |x|^{2}|y|^{2} +\langle x,y \rangle^{2} = - \sum_{j<k}  (x_{j} y_{k} - x_{k}y_{j})^{2},
\end{align*}
from which we observe that $(\ux \wedge \uy )^{2}$ is real-valued. Moreover, this allows us to estimate
\begin{equation}
\left| \frac{x_{j} y_{k} - x_{k}y_{j}}{\sqrt{|x|^{2}|y|^{2} -\langle x,y \rangle^{2}}} \right| \leq 1, \qquad \forall x, y \in \mR^{m}.
\label{estimateCterm}
\end{equation}

Finally, we introduce a basis $\{ \psi_{j,k,\ell}\}$ for the space $\cS(\mR^{m}) \otimes \cC l_{0, m}$, where 
$\cS(\mR^m)$ denotes the Schwartz space.  This basis is defined by 
\begin{align} \label{basis}
\begin{split}
\psi_{2j,k,\ell} &:= L_{j}^{\frac{m}{2}+k-1}(|x|^{2}) M_{k}^{(\ell)} e^{-|x|^{2}/2},\\
\psi_{2j+1,k,\ell} &:= L_{j}^{\frac{m}{2}+k}(|x|^{2}) \ux M_{k}^{(\ell)} e^{-|x|^{2}/2},
\end{split}
\end{align}
where $j,k \in \mN$, $\{M_{k}^{(\ell)} \in \cM_{k}: \ell= 1, \ldots, \dim \cM_{k}\}$ is a basis for $\cM_{k}$,
and $L_{j}^{\alpha}$ are again the Laguerre polynomials. The set $\{ \psi_{j,k,\ell}\}$ forms a basis of 
$\cS(\mR^{m}) \otimes \cC l_{0, m}$ (as can be seen from (\ref{Fischer})). Originally, these functions were called the Clifford-Hermite functions, see \cite{MR926831}. They have been generalized to different geometries such as the superspace case \cite{MR2371128} and the case of hermitean Clifford analysis \cite{MR2454729}.

\subsection{Clifford-Fourier transform}
\label{CFTsection}

The Clifford-Fourier transform (CFT) was first defined in \cite{MR2190678}. Its operator exponential definition is given by (following the normalization given in \cite{AIEP})
\begin{equation}
\cF_{\pm} =e^{ \frac{i \pi m}{4}} e^{\frac{i \pi}{4}(\Delta - |x|^{2} \mp 2\Gamma_x)} = e^{ \frac{i \pi m}{4}} e^{\mp \frac{i \pi}{2}\Gamma_x  }e^{\frac{i \pi}{4}(\Delta - |x|^{2})}.
\label{CFTF}
\end{equation}
The second equality follows because $\Gamma_x$ commutes with $\Delta$ and $|x|^{2}$. The motivation behind this definition was to find a pair of transforms $\cF_+$, $\cF_-$ that factorize the square of the Fourier transform for spinor- or Clifford algebra-valued functions. Schematically this is obtained in the following way:
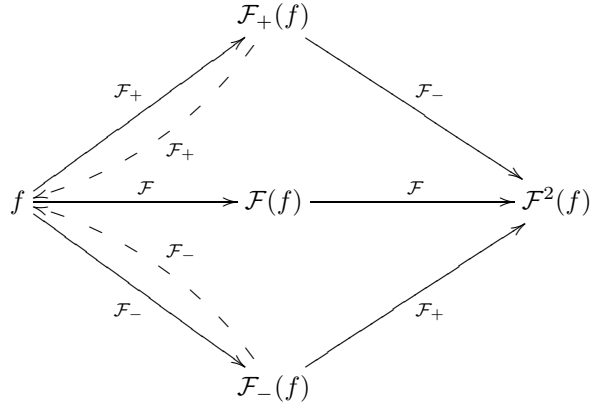
\[
\xymatrix{
&&\cF_+(f)\ar[rrdd]^{\cF_-} \ar@{-->}@/^1pc/[ddll]^{\cF_+}&&\\
&&\\
f \ar[rr]^{\cF} \ar[rruu]^{\cF_+} \ar[rrdd]_{\cF_-}&\qquad&\cF(f) \ar[rr]^{\cF}&\qquad&\cF^2(f)\\
&&&&\\
&&\cF_-(f)\ar[rruu]_{\cF_+} \ar@{-->}@/_1pc/[uull]_{\cF_-}&&
}
\]
which follows immediately from the exponential operator definition. A formal version of this result is given in the subsequent Theorem \ref{EigenfunctionsKernel}. Note that the dashed line is only true in case of even dimension, see again Theorem \ref{EigenfunctionsKernel}. In general, the inverse of the CFT is obtained as follows
\[
\cF_{\pm}^{-1} =e^{-  \frac{i \pi m}{4}} e^{-\frac{i \pi}{4}(\Delta - |x|^{2} \mp 2\Gamma_x )}.
\]

Combining formulas \textbf{F1} and \textbf{F3} for the classical Fourier transform with the definition of the Clifford-Fourier transform (\ref{CFTF}), we are immediately led to consider the two following integral transforms:
\begin{definition} \label{def:FC}
On the Schwartz class of functions $\cS(\mR^m) \otimes \cC l_{0, m}$, we define
\begin{align*}
\cF_{\pm}(f)(y) & : = (2 \pi)^{-\frac{m}{2}} \int_{\mR^{m}} K_\pm(x,y) f(x) \, dx,\\
\cF_{\pm}^{-1} (f)(y) & :=(2 \pi)^{-\frac{m}{2}} \int_{\mR^{m}} \widetilde{K_\pm}(x,y) f(x) \, dx,
\end{align*}
where
\begin{align*}
K_\pm(x,y)& := e^{\mp i \frac{\pi}{2}\Gamma_{y}} e^{-i \langle x,y \rangle},\\
\widetilde{K_\pm}(x,y)& := e^{ \pm i \frac{\pi}{2}\Gamma_{y}} e^{i \langle x,y \rangle}.
\end{align*}
\end{definition}

Note that at this point it is not yet clear that $\cF_{\pm}^{-1}$ is indeed the inverse operator for $\cF_\pm$. This will only be obtained in Theorem \ref{EigenfunctionsKernel}.

Explicit computation of the integral kernel $K_\pm(x,y)$ of the CFT is a hard problem. The so-called Clifford-Fourier kernel was first constructed in the case $m=2$ in \cite{MR2283868}. For higher even dimensions, a complicated iterative procedure for constructing the kernel was given in \cite{JFAA-Fourier-Bessel}, which could only be used practically in low dimensions. A breakthrough, leading to explicit formulas in all even dimensions, was obtained in \cite{DBXu}. In the rest of this subsection, we follow that paper.

\begin{remark}
It is important to note that these kernels are not symmetric, in the sense that $K(x,y) \neq K(y,x)$ (see e.g. Theorem \ref{seriesthm}
below). Hence, we adopt the convention that we always integrate over the first variable in the kernel.
\end{remark}

First step in determining the explicit expression for the kernel of the CFT is to obtain a series expansion of the type \textbf{F4}. This is the result of the following theorem, where we introduce new variables $z = |x||y|$ and $w = \langle \xi,\eta \rangle$ to simplify notations and substitute $m = 2 \lambda + 2$. This result was obtained using the projection operators in formula (\ref{projGamma}) and the action of the Gamma operator given in (\ref{gammaeig1}).

\begin{theorem}
The kernel of the Clifford-Fourier transform is given by $K_-(x,y) =  A_{\lambda} + B_{\lambda} + \left(\ux \wedge \uy \right) C_{\lambda}$
with
\begin{align*}
A_{\lambda}(w,z) =& \,2^{\lambda-1} \Gamma(\lambda+1)\sum_{k=0}^{\infty}  (i^{2 \lambda + 2} + (-1)^{k})  z^{-\lambda} J_{k+ \lambda}(z) \;  C_{k}^{\lambda}(w),\\
B_{\lambda}(w,z) = & \, - 2^{\lambda-1} \Gamma(\lambda)\sum_{k=0}^{\infty}(k+ \lambda)  (i^{2 \lambda + 2} - (-1)^{k}) z^{-\lambda} J_{k+ \lambda}(z) \; C_{k}^{\lambda}(w),\\
C_{\lambda}(w,z) = & \, -  2^{\lambda-1} \Gamma(\lambda)\sum_{k=0}^{\infty} (i^{2 \lambda + 2} + (-1)^{k}) z^{-\lambda-1} J_{k+ \lambda}(z) \;  \left(\frac{d}{dw}C_{k}^{\lambda}\right) (w),
\end{align*}
where $z = |x||y|$ and $w = \langle \xi,\eta \rangle$.
\label{seriesthm}
\end{theorem}

The functions $A_{\lambda}$, $B_{\lambda}$ and $C_{\lambda}$ satisfy nice recursive relations. They are given in the following lemma, which follows immediately from the formula $\frac{d}{d w} C^{\lambda}_{k}(w) = 2 \lambda  C^{\lambda+1}_{k-1}(w)$.
\begin{lemma} \label{RecursionsProps}
For $m > 2$, or equivalently, $\lambda >0$, one has
\begin{align*}
A_{\lambda}(w,z) =& \, - \frac{\lambda}{\lambda -1} \frac{1}{z} \partial_{w}A_{\lambda-1}(w,z),\\
B_{\lambda}(w,z) =& \, -  \frac{1}{z} \partial_{w}B_{\lambda-1}(w,z),\\
C_{\lambda}(w,z) =& \, -\frac{1}{\lambda z} \partial_{w} A_{\lambda}(w,z).
\end{align*}
\end{lemma}
Because of this lemma, which yields a recursion of size 2 on the dimension, it clearly suffices to explicitly compute the kernel in dimension 2 and 3. As we will see, this can be done in dimension 2. The case of dimension 3 (and consequently all odd dimensions) is still open.

We also need to relate the kernels for $\cF_+$ and the inverse transforms to the kernel of $\cF_-$. After establishing a similar series expansion, one obtains the following proposition
\begin{proposition}
\label{kernelsrels}
For $x, y \in \mR^m$, one has
\begin{align*}
K_+(x,y) &= \left(K_-(x,-y)\right)^c,\\
\widetilde{K_\pm}(x,y) &= \left(K_\pm(x,y)\right)^c
\end{align*}
where $c$ denotes complex conjugation.
In particular, in the case $m$ even, $K_-(x,y)$ is real-valued and the complex conjugation can be omitted.
\end{proposition}

As a consequence of this proposition, we immediately observe that in even dimension $\cF_{\pm}$ is its own inverse, as already indicated on the scheme with the dashed line.

\subsubsection{Clifford-Fourier transform in even dimension}

Let us now study the CFT in even dimension. We will start by giving the explicit formula in dimension 2. This was first obtained in \cite{MR2283868} using Clifford analysis techniques, namely projection operators. Subsequently, the same result was obtained in \cite{DBXu} by explicitly computing the series $A_{\lambda}$, $B_{\lambda}$, $C_{\lambda}$ for $\lambda =0$. The result is summarized in the following theorem.

\begin{theorem}
\label{KernelDim2}
The kernel of the Clifford-Fourier transform in dimension 2 is given by
\begin{align*}
K_-(x,y) &= \cos t + \left(\ux \wedge \uy \right) \frac{\sin t}{t}\\
& = e^{\, \ux \wedge \uy}
\end{align*}
with $t= |\ux \wedge \uy| =   \sqrt{|x|^{2} |y|^{2}-\langle x, y  \rangle^2}$.
\end{theorem}

For general even dimension, the explicit formulas are obtained by combining Lemma \ref{RecursionsProps} with the result in dimension 2. Then by using induction one proves the following result.

\begin{theorem}  \label{EvenExplicit}
The kernel of the Clifford-Fourier transform in even dimension $m>2$ is given by
\begin{align*}
K_-(x,y) &= e^{i \frac{\pi}{2}\Gamma_{y}} e^{-i \langle x,y \rangle}\\
& = (-1)^{\frac{m}{2}} \left(\frac{\pi}{2} \right)^{\frac{1}{2}} \left(A_{(m-2)/2}^{*}(s,t) +B_{(m-2)/2}^{*}(s,t)+ (\ux \wedge \uy) \ C_{(m-2)/2}^{*}(s,t)\right)
\end{align*}
where $s=\langle x,y \rangle$ and $t= |\ux \wedge \uy| =   \sqrt{|x|^{2} |y|^{2}-s^2}$ and
\begin{align*}
A_{(m-2)/2}^{*}(s,t) &=\sum_{\ell=0}^{\left\lfloor \frac{m}{4}-\frac{3}{4}\right\rfloor}  s^{m/2-2-2\ell} \ \frac{1}{2^{\ell} \ell!} \ \frac{\Gamma \left( \frac{m}{2} \right) }{ \Gamma \left( \frac{m}{2}-2\ell -1 \right)} J^*_{(m-2\ell-3)/2}(t),\\
B_{(m-2)/2}^{*}(s,t)&= - \sum_{\ell=0}^{\left\lfloor \frac{m}{4}-\frac{1}{2}\right\rfloor} s^{m/2-1-2\ell} \ \frac{1}{2^{\ell} \ell!} \ \frac{\Gamma \left( \frac{m}{2} \right)}{\Gamma \left( \frac{m}{2}-2\ell \right)} J^*_{(m-2\ell-3)/2}(t),\\
C_{(m-2)/2}^{*}(s,t) &= - \sum_{\ell=0}^{\left\lfloor \frac{m}{4}-\frac{1}{2}\right\rfloor} s^{m/2-1-2\ell} \ \frac{1}{2^{\ell} \ell!} \ \frac{\Gamma \left( \frac{m}{2} \right)}{\Gamma \left( \frac{m}{2}-2\ell \right)}  J^*_{(m-2\ell-1)/2}(t)
\end{align*}
with $J^*_{\alpha}(t) = t^{-\alpha} J_{\alpha}(t)$.
\end{theorem}

Note that in this theorem $J^*_\alpha(t)$ equals, up to a constant, $\widetilde{J}_\alpha(t)$. In order to avoid confusion for readers looking at the original papers, the formulas have not been rewritten in terms of the (more common) $\widetilde{J}_\alpha(t)$, which is used in the other sections of the present review.

For odd dimension, the problem is again reduced to computing the kernel in dimension 3. However, this is still a difficult open problem and the techniques used to prove the case of dimension 2 do not generalize. The best result obtained so far is an integral expression, see Lemma 4.5 in \cite{DBXu}. Note that this expression does not allow for the computation of a bound on the kernel.

Let us now turn our attention to the analytic properties of the CFT. Similar to the classical case, the CFT satisfies some calculus rules, which translates to the following system of equations satisfied by the kernel:
\begin{align} \label{C-F system}
\begin{split}
\partial_{\uy} \lbrack K_{\mp}(x,y) \rbrack = \mp \ (\pm i)^m \ K_{\pm}(x,y) \ \ux\\
\lbrack K_{\pm}(x,y) \rbrack \partial_{\ux} = \pm (\mp i)^m \ \uy \ K_{\mp}(x,y),
\end{split}
\end{align}
where 
\[
\lbrack K_{\pm}(x,y) \rbrack \partial_{\ux}  = \sum_{i=1}^{m} \left(\partial_{x_{i}}K_{\pm}(x,y) \right) e_{i}
\]
denotes the action of the Dirac operator on the right. The system of PDEs (\ref{C-F system}) should be compared with the formulation \textbf{F2} of the classical Fourier transform. We will come back to this Clifford-Fourier system in Section \ref{CFTclasssec}.

Again in the even dimensional case, we can obtain a bound for the kernel.

\begin{lemma} \label{lem:bound}
Let $m$ be even. For $ x, y  \in \mR^m$, there exists a constant $c$ such that
\begin{align*}
|A_{(m-2)/2}^{*}(s,t) + B_{(m-2)/2}^{*}(s,t)| & \leq  c (1+|x|)^{(m-2)/2}(1+|y|)^{(m-2)/2},  \\
|(x_{j} y_{k} - x_{k}y_{j}) C_{(m-2)/2}^{*}(s,t)| & \leq  c (1+|x|)^{(m-2)/2}(1+|y|)^{(m-2)/2}, \qquad
 j \neq k. 
\end{align*}
\end{lemma}
As an immediate consequence of this lemma, the domain in the definition of the Clifford-Fourier transform can be specified. Let us define a class of functions
$$
    B_{(m-2)/2}(\mR^m) : = \left\{ f\in L^1(\mR^m): \int_{\mR^m} (1+|y|)^{(m-2)/2} | f(y)| dy < \infty \right\}.
$$

\begin{theorem} \label{CFdomain}
Let $m$ be an even integer. The Clifford-Fourier transform is well-defined on $B_{(m-2)/2}(\mR^m) \otimes \cC l_{0, m}$. In particular, for 
$f \in B_{(m-2)/2}(\mR^m) \otimes \cC l_{0, m}$, $\cF_\pm f$ is a continuous function.
\end{theorem}

For $m$ being even we can now establish the inversion formula for Schwartz class functions. First we state the operational calculus, based on formula (\ref{C-F system}). 

\begin{lemma} \label{DiffTransform1}
Let $m$ be even and $f \in \cS(\mR^{m})\otimes \cC l_{0, m}$. Then  
\begin{align*}
\cF_{\pm} \left(\ux \, f \right) & = \mp (-1)^{m/2} \upy \cF_{\mp} \left( f \right), \\
\cF_{\pm} \left(\upx f \right) & = \mp (-1)^{m/2}  \uy \cF_{\mp} \left( f \right).
\end{align*} 
\end{lemma}

Using this lemma, one can show that the kernel indeed defines a continuous integral operator. The proof of this theorem is rather technical and long, as it requires estimates of suitable seminorms on the Schwartz space.
\begin{theorem} \label{CFschwartz} 
Let $m$ be even. Then $\cF_\pm$ is a continuous operator on $\cS(\mR^m) \otimes \cC l_{0, m}$.
\end{theorem}

Finally, we state a theorem that gives the eigenvalues of the CFT. Using the continuity and the density of the basis $\{\psi_{j,k,\ell}\}$, this also yields the inversion in Schwartz space.

\begin{theorem} \label{EigenfunctionsKernel}
For the basis $\{\psi_{j,k,\ell}\}$ of  $\cS(\mR^{m}) \otimes \cC l_{0, m}$ , one has
\begin{align*}
\cF_{\pm}(\psi_{2j,k,\ell}) &= (-1)^{j+k} (\mp 1)^{k} \psi_{2j,k,\ell},\\
\cF_{\pm} (\psi_{2j+1,k,\ell}) &= i^{m} (-1)^{j+1} (\mp 1 )^{k+m-1} \psi_{2j+1,k,\ell}.
\end{align*}
In particular, the action of $\cF_\pm$ coincides with the operator $e^{ \frac{i \pi m}{4}} e^{\mp \frac{i \pi}{2}\Gamma_x  }e^{\frac{i \pi}{4}(\Delta - |x|^{2})}$ when restricted to the basis $\{ \psi_{j,k,\ell}\}$ and 
\begin{equation}\label{inversionCF}
\cF_{\pm}^{-1} \cF_{\pm} = id 
\end{equation}
on the basis $\{\psi_{j,k,\ell}\}$, with $\cF_{\pm}^{-1}$ as in definition \ref{def:FC}. Moreover, when $m$ is even, \eqref{inversionCF} holds for all $f \in \cS(\mR^m) \otimes \cC l_{0, m}$ and one has that $\cF_{\pm}^{-1} = \cF_{\pm}$.
\end{theorem}

To conclude this section, let us remark that it is also possible to prove a more general inversion theorem. The proof is very long (see \cite{DBXu}), as it requires the introduction of a translation operator and convolution theorem for the Clifford-Fourier transform. The final result is given in the following theorem.

\begin{theorem} 
Let $m$ be even. If $f \in B_{(m-2)/2}(\mR^m) \otimes \cC l_{0, m}$ and $ \cF_- f \in B_{(m-2)/2}(\mR^m) \otimes \cC l_{0, m}$, and if 
$$
   g(x) = (2\pi)^{-m/2} \int_{\mR^m}  K_-(\xi,x) \cF_- f(\xi) d \xi, \qquad  x \in \mR^m, 
$$
then $g \in C(\mR^m)$ and $f(x) = g(x)$ a.e.
\end{theorem}

\subsubsection{A fractional version of the Clifford-Fourier transform}

To get a better insight in the role played by the two different exponential factors in the definition of the CFT, it is interesting to introduce a fractional version of this transform. This fractional CFT was defined in \cite{DBDSFr} by
\[
\cF_{\alpha, \b} =   e^{ \frac{i \alpha m}{2}} e^{i\b \Gamma_x}e^{ \frac{i \alpha}{2}(\Delta - |x|^{2})}
\]
where $\alpha, \b \in [-\pi,\pi]$. The CFT studied in the previous section is then reobtained via $\cF_{\pm} = \cF_{ \pi/2, \mp \pi/2 }$.

Observe also that we immediately have
\[
\cF_{\alpha, \b} \circ \cF_{\alpha, - \b} = \cF_{\alpha, -\b} \circ \cF_{\alpha,  \b} =  \cF_{\alpha}^{2} = \cF_{2 \alpha}
\]
with $\cF_{\alpha} =   \cF_{\a,0}= e^{ \frac{i \alpha m}{2}}e^{ \frac{i \alpha}{2}(\Delta - |x|^{2} )}
 $ the fractional version of the ordinary Fourier transform, introduced in section \ref{sec_frFT}.
 
Schematically, an overview of the action of the fractional CFT is given by
\[
\xymatrix{
&&\cF_{\a, \b}(f)\ar[rrdd]^{\cF_{-\a, -\b}} \ar[dd]_{\cF_{0,-\b}}&&\\
&&\\
f \ar[rr]^{\cF_{\a, 0}} \ar[rruu]^{\cF_{\a, \b}} &\qquad&\cF_{\a, 0}(f) \ar[rr]^{\cF_{-\a,0}}&\qquad&f
}
\]

Again the main problem is to find an integral expression for $\cF_{ \alpha, \b}$:
\[
\cF_{ \alpha,\b}(f)(y) =  \left(\pi (1- e^{-2i \alpha})\right)^{-m/2}\int_{\mathbb{R}^m} K_{\alpha,\b}(x,y) \ f(x) \ dx.
\]
In our computation, we need to put a few restrictions on the parameters $\alpha$ and $\beta$. We exclude the case where $\alpha =0$ or $\alpha =\pm \pi$, as then the transforms reduce to singular integral operators on the sphere or delta distributions (see \cite{DBDSFr}).

The series expansion of the kernel was obtained in \cite{DBDSFr}. Note that the result stays fairly close to that for the CFT.
\begin{theorem}
\label{FracSeries}
The fractional Clifford-Fourier transform $\cF_{ \alpha, \b} =  e^{ \frac{i \alpha m}{2}} e^{ i \b \Gamma_x} e^{ \frac{i \alpha}{2}(\Delta - |x|^{2})}$ is given by the integral transform
\[
\left(\pi (1- e^{-2i \alpha})\right)^{-m/2}\int_{\mathbb{R}^m} K_{\alpha,\b}(x,y) \ f(x) \ dx
\]
with integral kernel 
\[
K_{\alpha,\b}(x,y) =  \left(A_{\lambda}^{\alpha,\b} + B_{\lambda}^{\alpha,\b} + \left(\ux \wedge \uy \right) C_{\lambda}^{\alpha,\b}\right)e^{\frac{i}{2}( \cot \alpha) (|x|^2 + |y|^2)} 
\]
with
\begin{align*}
A_{\lambda}^{\alpha,\b}(w,\widetilde{z}) =& \,- 2^{\lambda-1} \Gamma(\lambda+1)\sum_{k=0}^{\infty} i^{-k}(e^{i \b (k+2 \lambda)} - e^{-i \b k}) \widetilde{z}^{-\lambda} J_{k+ \lambda}(\widetilde{z}) \;  C_{k}^{\lambda}(w),\\
B_{\lambda}^{\alpha,\b}(w,\widetilde{z}) = & \,  2^{\lambda-1} \Gamma(\lambda)\sum_{k=0}^{\infty}(k+ \lambda) i^{-k}(e^{i \b (k+2 \lambda)} +e^{-i \b k})\widetilde{z}^{-\lambda} J_{k+ \lambda}(\widetilde{z}) \; C_{k}^{\lambda}(w),\\
C_{\lambda}^{\alpha,\b}(w,\widetilde{z}) = & \,   \frac{2^{\lambda} \Gamma(\lambda+1)}{\sin{\alpha}}\sum_{k=1}^{\infty} i^{-k}(e^{i \b (k+2 \lambda)} - e^{-i \b k}) \widetilde{z}^{-\lambda-1} J_{k+ \lambda}(\widetilde{z}) \;  C_{k-1}^{\lambda+1} (w),
\end{align*}
where $\widetilde{z} = (|x||y|)/ \sin{\alpha}$, $w = \langle\xi,\eta \rangle$ and $\lambda = (m-2)/2$.
\end{theorem}

The kernel can again be computed explicitly in the case of even dimension, using similar recursion properties as for the CFT. Let us first discuss the case of dimension 2. This was obtained in \cite{DBDSC} for $\alpha=\beta$ and extended to arbitrary $\a$ and $\b$ in \cite{DBDSFr}.

\begin{theorem}
\label{KernelDim2Fr}
The kernel of the fractional Clifford-Fourier transform in dimension $m=2$ is given by
\begin{align*}
K_{\alpha,\b}(x,y) &= e^{\frac{i}{2}( \cot \alpha) (|x|^2 + |y|^2)}  e^{ i \b \Gamma_{y}} \left(  e^{-i \la x,y \ra / \sin \alpha}\right)\\
& = \left(\cos{ \left(t\frac{\sin{\b}}{\sin{\a}}\right)} + \left(\ux \wedge \uy \right) \frac{\sin{\left(t \frac{\sin{\b}}{\sin{\a}}\right)}}{t}\right)e^{-i \langle x,y \rangle\frac{\cos{\b}}{\sin{\a}}}e^{\frac{i}{2}( \cot \alpha) (|x|^2 + |y|^2)} 
\end{align*}
with $t= |\ux \wedge \uy| =   \sqrt{|x|^{2} |y|^{2}-\langle x, y  \rangle^2}$.
\end{theorem}

Now we can state the result for general even dimension. Note that the resulting kernel is a complicated expression, depending on both fractional parameters $\a$ and $\b$.

\begin{theorem}\label{EvenExplicitFr}
The kernel of the fractional Clifford-Fourier transform in even dimension $m>2$ is given by
\begin{align*}
K_{\alpha,\b}(x,y) &= e^{\frac{i}{2}( \cot \alpha) (|x|^2 + |y|^2)}  e^{ i \b \Gamma_{y}} \left(  e^{-i \la x,y \ra / \sin \alpha}\right)\\
& =  \left(\frac{\pi}{2} \right)^{\frac{1}{2}} e^{i \b \frac{m-2}{2}} (\cos{\b})^{\frac{m-2}{2}} e^{-i s^{\ast} \cot{\b}} e^{\frac{i}{2}( \cot \alpha) (|x|^2 + |y|^2)} \\
& \quad \times
 \left(A_{(m-2)/2}^{\alpha,\b, *}(s^{\ast},t^{\ast}) +B_{(m-2)/2}^{\alpha,\b, *}(s^{\ast},t^{\ast})+ (\ux \wedge \uy) \ C_{(m-2)/2}^{\alpha,\b, *}(s^{\ast},t^{\ast})\right) 
\end{align*}
where $s^{\ast}= \frac{\sin{\b}}{\sin{\alpha}} \langle x,y \rangle$ and $t^{\ast}= \frac{\sin{\b}}{\sin{\alpha}} |\ux \wedge \uy|$ and
\begin{align*}
A_{(m-2)/2}^{\alpha,\b,*}(s^{\ast},t^{\ast}) &= i \ \left( \frac{m-2}{2}\right) \tan{\b} \sum_{\ell=0}^{\frac{m}{2}-2} 
\binom{\frac{m}{2}-2}{\ell} (i \tan{\b})^{\ell} \ C_{\ell}^{*}(s^{\ast},t^{\ast}),\\
B_{(m-2)/2}^{\alpha,\b,*}(s^{\ast},t^{\ast})&= - \sum_{\ell=0}^{\frac{m}{2}-1} 
\binom{\frac{m}{2}-1}{\ell} (i \tan{\b})^{\ell} \ B_{\ell}^{*}(s^{\ast},t^{\ast}),\\
C_{(m-2)/2}^{\alpha,\b,*}(s^{\ast},t^{\ast}) &= -  \frac{\sin{\b}}{\sin{\alpha}} \sum_{\ell=0}^{\frac{m}{2}-1} 
\binom{\frac{m}{2}-1}{\ell} (i \tan{\b})^{\ell} \ C_{\ell}^{*}(s^{\ast},t^{\ast})
\end{align*}
with $B_{\ell}^{*}(s^{\ast},t^{\ast})$ and $C_{\ell}^{*}(s^{\ast},t^{\ast})$ defined in Theorem \ref{EvenExplicit}.
\end{theorem}

Most results for the CFT carry over to the fractional CFT. In particular, the kernel of the fractional CFT obeys a similar polynomial bound and the integral transform again yields a continuous map on the Schwartz space. For precise details, see \cite{DBDSFr}.

Putting
\[
K_{\alpha,\beta}(x,y) = \widehat{K}_{\alpha,\beta}(x,y) \  e^{\frac{i}{2}( \cot \alpha) (|x|^2 + |y|^2)},
\]
let us determine the system of PDEs satisfied by $\widehat{K}_{\alpha,\beta}$.
\begin{proposition}
\label{simpleSystem}
 For all $m$, $\widehat{K}_{\alpha,\beta}(x,y)$ satisfies
\begin{align*}
(i \sin{\alpha} \ \partial_{\uy}) \lbrack \widehat{K}_{\alpha,\b}(x,y) \rbrack &= e^{i\b(m-1)} \widehat{K}_{\alpha,-\b}(x,y) \ \ux,\\
\uy \widehat{K}_{\alpha,\b}(x,y)&= e^{i\b(m-1)}  \lbrack \widehat{K}_{\alpha,-\b}(x,y) \rbrack  (i \sin{\alpha} \ \partial_{\ux}).
\end{align*}
\end{proposition}

Again, this proposition should be compared with formulation \textbf{F2} of the ordinary FT. As a consequence, we obtain the calculus properties of the fractional CFT.

\begin{lemma} \label{DiffTransform2}
Let $m$ be even and $f \in \cS(\mR^{m}) \otimes \cC l_{0, m}$. Then  
\begin{align*}
\cF_{\alpha, -\b} \left((\ux\cos{\a}-i \sin{\a} \upx) \, f \right) & = e^{-i \b(m-1)} \uy \cF_{\alpha, \b} \left( f \right), \\
\cF_{\alpha, -\b} \left((\cos{\a} \upx - i \ux \sin{\a}) f \right) & = e^{-i \b(m-1)} \upy \cF_{\alpha, \b} \left( f \right).
\end{align*} 
\end{lemma}

Let us now state the final result concerning the fractional CFT, that explains our previous scheme.

\begin{theorem}
\label{InvTheoremS}
For the basis $\{\psi_{j,k,\ell}\}$ of  $\cS(\mR^{m}) \otimes \cC l_{0, m}$ , one has
\begin{align*}
\cF_{\alpha,\b}\left( \psi_{2j,k,\ell} \right) &= e^{-i\alpha(2j+k)} e^{-i\b k} \ \psi_{2j,k,\ell},\\
\cF_{\alpha,\b} \left( \psi_{2j+1,k,\ell} \right) &= e^{-i\alpha(2j+1+k)} e^{i\b (k+m-1)} \ \psi_{2j+1,k,\ell}.
\end{align*}
In particular, the action of $\cF_{\alpha,\b}$ coincides with the operator $e^{ \frac{i \alpha m}{2}} e^{i\b \Gamma_x}e^{ \frac{i \alpha}{2}(\Delta - |x|^{2})}
$ when restricted to the basis $\{ \psi_{j,k,\ell}\}$ and
\begin{equation}\label{inversion}
\cF_{\alpha,\b} \cF_{-\alpha,-\b}= id 
\end{equation}
on the basis $\{\psi_{j,k,\ell}\}$. Moreover, when $m$ is even, \eqref{inversion} holds for all $f \in \cS(\mR^m) \otimes \cC l_{0, m}$.
\end{theorem}

\subsubsection{An entire class of Clifford-Fourier transforms}
\label{CFTclasssec}

In this subsection we again restrict ourselves to the case of even dimension. Similar computations have been performed in the odd case, see \cite{DBNS}, Section 5 for a detailed exposition.

The aim of this section is to discuss solutions of the Clifford-Fourier system (\ref{C-F system}) in even dimension:
\begin{align}\label{system}
\begin{split}
\partial_{\uy} \lbrack K_+(x,y) \rbrack  =  a \ K_-(x,y) \ \ux\\
\lbrack K_+(x,y) \rbrack \partial_{\ux}  =  a \ \uy \ K_-(x,y)
\end{split}
\end{align}
with $ a=(-1)^{m/2}$ and $K_-(x,y) = K_+(x,-y)$, see Proposition \ref{kernelsrels}. Note that the similar system \textbf{F2} for the ordinary Fourier transform has a unique solution when we impose a suitable initial condition. It turns out that this is no longer the case for the Clifford-Fourier system. Indeed, not only the kernel of the CFT but also the kernel of the so-called Fourier-Bessel transform \cite{FourierBessel} satisfies this system. It is hence an obvious problem to find all solutions of this system, which was done in \cite{DBNS}. Let us describe the procedure that was followed.

As the even dimensional CFT is real-valued, we look for solutions $K_+(x,y)$ with real-valued components. Inspired by the explicit expression in Theorem \ref{EvenExplicit} and Proposition \ref{kernelsrels}, the aim is to find all solutions of the form:
\begin{align*}
K_+(x,y) &= f(s,t) + (\ux \wedge \uy) \ g(s,t)\\
K_-(x,y) &=  f(-s,t) - (\ux \wedge \uy) \ g(-s,t)
\end{align*}
with $s = \langle x,y \rangle$, $t=|\ux \wedge \uy|$ and $f$ and $g$ real-valued functions. Substituting these expressions in (\ref{system}) leads to the following systems of PDEs for $f$ and $g$
\begin{align}\label{fg}
\begin{split}
\partial_s \lbrack f(s,t) \rbrack + t \partial_t \lbrack g(s,t)\rbrack + (m-1) \ g(s,t)  &=  a \ f(-s,t)\\
\partial_s \lbrack g(s,t) \rbrack - \frac{1}{t} \partial_t \lbrack f(s,t) \rbrack  &=  a \ g(-s,t).
\end{split}
\end{align}
We want to find solutions of the system (\ref{fg}) which are as close as possible to the kernel of the Clifford-Fourier transform given in Theorem \ref{EvenExplicit}. Therefore, we propose to find all solutions of the form
\[
f(s,t)  =  \sum_{j=0}^k s^{k-j} \ f_j(t), \qquad g(s,t)  =  \sum_{j=0}^k s^{k-j} \ g_j(t)
\]
with $k \in \mathbb{N}$ a parameter. In other words, we want the solution to be polynomial in $s$, but do not prescribe the behavior of the $t$ variable.

As shown by tedious computations, it turns out that there are $m-1$ such solutions. They are given explicitly by the following formulas:
\begin{equation}\label{kernel even m}
K^{j}_{+ , m}(x,y)  =  \tilde{f}_m^j(s,t) +  \hat{f}_m^j(s,t) + (\ux \wedge \uy) \ g_m^j(s,t), \qquad j=0,1,2, \ldots , m-2
\end{equation}
with
\begin{align*}
\tilde{f}_m^j(s,t) &=  -  \sqrt{\frac{\pi}{2}} \ \sum_{\ell =0}^{\left\lfloor  \frac{j-1}{2} \right\rfloor} s^{j-1-2 \ell} \ \frac{1}{2^{\ell} \ell!} \frac{\Gamma(j+1)}{\Gamma(j-2\ell)} \ J^*_{(m-2\ell-3)/2}(t), \qquad j \geq 1\\
\hat{f}_m^j (s,t) &=  (-1)^{m/2+j} \   \sqrt{\frac{\pi}{2}} \ \sum_{\ell =0}^{\left\lfloor  \frac{j}{2} \right\rfloor} s^{j-2 \ell} \ \frac{1}{2^{\ell} \ell!} \frac{\Gamma(j+1)}{\Gamma(j+1-2\ell)} \ J^*_{(m-2\ell-3)/2}(t) , \qquad j \geq 0\\
g_m^j(s,t) &=   \sqrt{\frac{\pi}{2}} \ \sum_{\ell =0}^{\left\lfloor  \frac{j}{2} \right\rfloor} s^{j-2 \ell} \ \frac{1}{2^{\ell} \ell!} \frac{\Gamma(j+1)}{\Gamma(j+1-2\ell)} \ J^*_{(m-2\ell-1)/2}(t) , \qquad  j \geq 0.
\end{align*}
We also put $\tilde{f}_m^0(s,t)=0$.

Note that
\begin{itemize}
\item the Fourier-Bessel kernel \cite{FourierBessel} is obtained for $j=0$. Hence we put $K_{+,m}^{0}(x,y) = K_{+,m}^{\mathrm{Bessel}} (x,y)$. 
\item the even dimensional Clifford-Fourier kernel (see Theorem \ref{EvenExplicit}) is obtained, up to a minus sign, for $j = \frac{m}{2}-1$. Hence we denote $K_{+,m}^{m/2-1}(x,y) = K_{+,m}^{\mathrm{CF}} (x,y)$.
\item as $m$ is even, the solution is given in terms of Bessel functions of order $n+\frac{1}{2}$ with $n \in \mathbb{N}$. In odd dimensions, the Bessel functions obtained are of integer order, see \cite{DBNS}.
\end{itemize}
It is possible to arrange all the kernels as in the scheme below. The middle line in the diagram corresponds to the Clifford-Fourier kernel, the lower diagonal with the Fourier-Bessel kernel.
 At each step in the dimension, two new kernels appear ($K_{+,m}^0$ and $K_{+,m}^{m-2}$) corresponding to the Fourier-Bessel kernel and its inverse (as will follow from Theorem \ref{TFonS}). The other kernels at a given step in the dimension ($K_{+,m}^j$, $j=1,2,\ldots,m-3$) follow from the previous dimension $m-2$ by a suitable action of a differential operator, see \cite{DBNS}, Proposition 4.1. 
\[
\xymatrix@=12pt{m=2&m=4&m=6&m=8\\
&&&K_{+,8}^{6}\\
&&K_{+,6}^{4}\ar@/^/[ur] \ar[r]^-{z^{-1}\partial_{w}}&K_{+,8}^{5}\\
&K_{+,4}^{2} \ar@/^/[ur]\ar[r]^-{z^{-1}\partial_{w}}&K_{+,6}^{3}\ar[r]^-{z^{-1}\partial_{w}}&K_{+,8}^{4}\\
K_{+,2}^{0} \ar@/^/[ur] \ar@/_/[dr] \ar[r]^-{z^{-1}\partial_{w}}&K_{+,4}^{1} \ar[r]^-{z^{-1}\partial_{w}}&K_{+,6}^{2} \ar[r]^-{z^{-1}\partial_{w}}&K_{+,8}^{3}\\
&K_{+,4}^{0} \ar@/_/[dr] \ar[r]^-{z^{-1}\partial_{w}}&K_{+,6}^{1}\ar[r]^-{z^{-1}\partial_{w}}&K_{+,8}^{2}\\
&&K_{+,6}^{0} \ar@/_/[dr] \ar[r]^-{z^{-1}\partial_{w}}&K_{+,8}^{1}\\
&&&K_{+,8}^{0}
}
\]

In order to derive properties for these new kernels, we need to determine series expansions in terms of Bessel functions and Gegenbauer polynomials as in \textbf{F4}. This is the subject of the following theorem.

\begin{theorem} 
\label{SeriesEven}
The following series expansions hold:\vspace{0,2cm}\\
\emph{\underline{Case 1}: $j$ even ($j=0,2,\ldots,m-4,m-2$)}
\begin{align*}
\tilde{f}_m^j(w,z) &=  - j \ \left( \frac{m}{2}-2 \right)! \ 2^{m/2-2}  \  \sum_{k=0}^{\infty} (4k+m) \ \frac{(2k+j-1)!!}{(2k+m-j-1)!!} z^{-m/2+1} \ J_{2k+m/2}(z) \ C_{2k+1}^{m/2-1}(w)\\
 \hat{f}_m^j(w,z) &=  (-1)^{m/2} \ \left( \frac{m}{2}-2 \right)! \ 2^{m/2-1} \ \sum_{k=0}^{\infty} \left( 2k + \frac{m}{2}-1 \right) \ \frac{(2k+j-1)!!}{(2k-j+m-3)!!} z^{1-m/2} \ J_{2k+m/2-1}(z) \ C_{2k}^{m/2-1}(w)\\
g_m^j(w,z) &=   \left( \frac{m}{2}-1 \right)! \ 2^{m/2-1} \ \sum_{k=0}^{\infty} (4k+m) \ \frac{(2k+j-1)!!}{(2k+m-j-1)!!}  z^{-m/2} \ J_{2k+m/2}(z) \ C_{2k}^{m/2}(w)
\end{align*}
\emph{\underline{Case 2}: $j$ odd ($j=1,3,\ldots, m-5, m-3$)}
\begin{align*}
\tilde{f}_m^j(w,z) &= - j\ \left( \frac{m}{2}-2 \right)! \ 2^{m/2-2}  \  \sum_{k=0}^{\infty} (4k+m-2) \ \frac{(2k+j-2)!!}{(2k+m-j-2)!!}  z^{-m/2+1} \ J_{2k+m/2-1}(z) \ C_{2k}^{m/2-1}(w)\\
\hat{f}_m^j(w,z) &= (-1)^{m/2+1} \  \left( \frac{m}{2}-2\right)! \ 2^{m/2-1} \ \sum_{k=0}^{\infty} \left( 2k + \frac{m}{2} \right) \ \frac{(2k+j)!!}{(2k+m-j-2)!!}  z^{1-m/2} \ J_{2k+m/2}(z) \ C_{2k+1}^{m/2-1}(w)\\
g_m^j(w,z) &=   \left( \frac{m}{2}-1 \right)! \ 2^{m/2-1} \ \sum_{k=0}^{\infty}(4k+m+2) \  \frac{(2k+j)!!}{(2k+m-j)!!}   z^{-m/2} \ J_{2k+m/2+1}(z) \ C_{2k+1}^{m/2}(w) .
\end{align*}
\end{theorem}

We can now calculate the action of the new Clifford-Fourier transforms, defined by
\begin{displaymath}
\mathcal{F}^{j}_{+,m} \left( f \right) (y) = \frac{1}{(2\pi)^{m/2}} \ \int_{\mathbb{R}^m} K^{j}_{+,m}(x,y) \ f(x) \ dx, \qquad j=0,1,2,\ldots,m-2
\end{displaymath}
on the basis $\lbrace \psi_{p,k,\ell} \rbrace$. This yields the following result:

\begin{theorem}
\label{EigValsEven}
In case of $m$ even, the Clifford-Fourier transforms $\mathcal{F}^{j}_{+,m}$ act as follows on the basis $\lbrace \psi_{p,k,\ell} \rbrace$ of ${\mathcal S}(\mathbb{R}^m) \otimes \cC l_{0, m}$:\vspace{0,2cm}\\
\emph{\underline{Case 1}: $j+k$ even}
\begin{align*}
\mathcal{F}^{j}_{+,m} \left( \psi_{2p,k,\ell} \right) (y) &= (-1)^{m/2} (-1)^j \ \frac{(k+j-1)!!}{(k+m-j-3)!!} \ (-1)^p \ \psi_{2p,k,\ell}(y)\\
\mathcal{F}^{j}_{+,m} \left( \psi_{2p+1,k,\ell} \right) (y) &=   \frac{(k+j-1)!!}{(k+m-j-3)!!} \ (-1)^p \ \psi_{2p+1,k,\ell}(y)
\end{align*}

\noindent
\emph{\underline{Case 2}: $j+k$ odd}
\begin{align*}
\mathcal{F}^{j}_{+,m} \left( \psi_{2p,k,\ell} \right) (y) &=   \frac{(k+j)!!}{(k+m-j-2)!!} \ (-1)^{p+1} \ \psi_{2p,k,\ell}(y)\\
\mathcal{F}^{j}_{+,m} \left( \psi_{2p+1,k,\ell} \right) (y) &=  (-1)^{m/2} (-1)^j \  \frac{(k+j)!!}{(k+m-j-2)!!} \ (-1)^{p} \ \psi_{2p+1,k,\ell}(y).
\end{align*}
\end{theorem}

\begin{remark}
Putting $j=0$, we indeed obtain the eigenvalue equations of the Fourier-Bessel transform, while for $j=\frac{m}{2}-1$ the ones of the Clifford-Fourier transform appear (see Theorem \ref{EigenfunctionsKernel}). The result for the Fourier-Bessel transform was obtained earlier in \cite{FourierBessel} using complicated integral identities for special functions. 
\end{remark}

Again, similar bounds for these kernels exist as for the Clifford-Fourier transform.
\begin{lemma}
\label{bounds}
Let $m$ be even and $j= 0, \ldots, m-2$. For $ x, y  \in \mR^m$, there exists a constant $c$ such that
\begin{align*}
|\tilde{f}_m^j(s,t)  + \hat{f}_m^j(s,t) | &\leq  c (1+|x|)^{j}(1+|y|)^{j},  \\
|(x_{k} y_{\ell} - x_{\ell}y_{k}) g_m^j(s,t)| &\leq  c (1+|x|)^{j}(1+|y|)^{j}, \qquad
 k \neq \ell. 
\end{align*}
\end{lemma}

As an immediate consequence of Lemma \ref{bounds}, we can now specify the domain in the definition of the new class of Fourier transforms. Define the following function spaces, for $j = 1, \ldots, m-2$,
$$
B_{j}(\mR^m) : = \left\{ f\in L^{1}(\mR^m): \int_{\mR^m} (1+|y|)^{j} | f(y)| \ dy < \infty \right\}.
$$
Note that for $j=0$, $B_{0}(\mR^m) =L^{1}(\mR^m)$.
Then, in the spirit of formulation $\textbf{F1}$ of the ordinary Fourier transform, we have the following theorem.

\begin{theorem} \label{CFdomain2}
The integral transform  $\mathcal{F}^{j}_{+,m}$ is well-defined on $B_{j}(\mR^m) \otimes \cC l_{0, m}$. In particular, for $f \in B_{j}(\mR^m) \otimes \cC l_{0, m}$, $\mathcal{F}^{j}_{+,m}( f )$ is a continuous function.
\end{theorem}

This statement can be made more precise when considering Schwartz functions. In particular, it turns out that the inverse transform is found within the same class of transforms. This is a quite surprising result.

\begin{theorem}
\label{TFonS}
Let $j= 0, \ldots, m-2$. The integral transforms $\mathcal{F}^{j}_{+,m}$ define continuous operators mapping $\mathcal{S}(\mathbb{R}^m) \otimes \cC l_{0, m}$ to $\mathcal{S}(\mathbb{R}^m) \otimes \cC l_{0, m}$.

When $m$ is even, the inverse of each transform $\mathcal{F}^{j}_{+,m}$ is given by $\mathcal{F}^{m-2-j}_{+,m}$, i.e.
\begin{equation}
\label{invCFTclass}
\mathcal{F}^{j}_{+,m} \mathcal{F}^{m-2-j}_{+,m} = \mathcal{F}^{m-2-j}_{+,m} \mathcal{F}^{j}_{+,m} = id.
\end{equation}
In particular, when $j = (m-2)/2$, the transform reduces to the Clifford-Fourier transform, satisfying
\[
\mathcal{F}^{(m-2)/2}_{+,m} \mathcal{F}^{(m-2)/2}_{+,m} = id
\]
and the kernel is also given by
\begin{equation*}
\label{CFexpr}
K^{(m-2)/2}_{+,m}(x,y) =  - e^{- \frac{i \pi}{2}\Gamma_{y}  } \left( e^{-i \la x, y  \ra} \right).
\end{equation*}
\end{theorem}

In the following theorem we discuss the extension of the transforms $\mathcal{F}^{j}_{+,m}$ to $L^2(\mathbb{R}^m) \otimes \cC l_{0, m}$.

\begin{theorem}
The transform $\mathcal{F}^{j}_{+,m}$ extends from $\mathcal{S}(\mathbb{R}^m) \otimes \cC l_{0, m}$  to a continuous map on $L^2(\mathbb{R}^m) \otimes \cC l_{0, m}$ for all $j  \leq (m-2)/2$, but not for $j > (m-2)/2$. 

In particular, only when $m$ is even and $j = (m-2)/2$, the transform $\mathcal{F}^{(m-2)/2}_{+,m}$ is unitary, i.e.
\[
||\mathcal{F}^{(m-2)/2}_{+,m} (f) || = ||f||
\]
for all $f \in L^2(\mathbb{R}^m) \otimes \cC l_{0, m}$.
\end{theorem}

We can now also introduce the transforms $\mathcal{F}^{j}_{-,m}$ as
\[
\mathcal{F}^{j}_{-,m} (f) (y) =  \frac{1}{(2 \pi)^{m/2}} \ \int_{\mathbb{R}^m} K^{j}_{-,m}(x,y) \ f(x) \ dx
\]
with $K^{j}_{-,m}(x,y) = \left( K^{j}_{+,m}(x,-y) \right)^c$.  Note that $\mathcal{F}^{j}_{-,m} (f) (y) = \mathcal{F}^{j}_{+,m}(f) (-y)$, when $m$ is even.

Then we obtain the following proposition
\begin{proposition} \label{DiffTransform3}
Let $f \in \cS(\mR^{m})\otimes \cC l_{0, m}$ and $j=0,\ldots,m-2$. Then one has
\begin{align*}
\cF_{\pm,m}^j \left( \ux \, f \right) & = \mp \ (\mp i)^m \  \upy \left( \cF_{\mp,m}^j (  f ) \right)\\
\cF_{\pm,m}^j \left( \upx  f   \right) & = \mp  \ (\mp i)^m \ \uy \ \cF_{\mp,m}^j \left( f \right).
\end{align*} 
\end{proposition}
Note that the formulas in this proposition are independent of $j$, as expected.

Finally, let us summarize the action of the transforms in the CFT class for even dimension with a scheme:
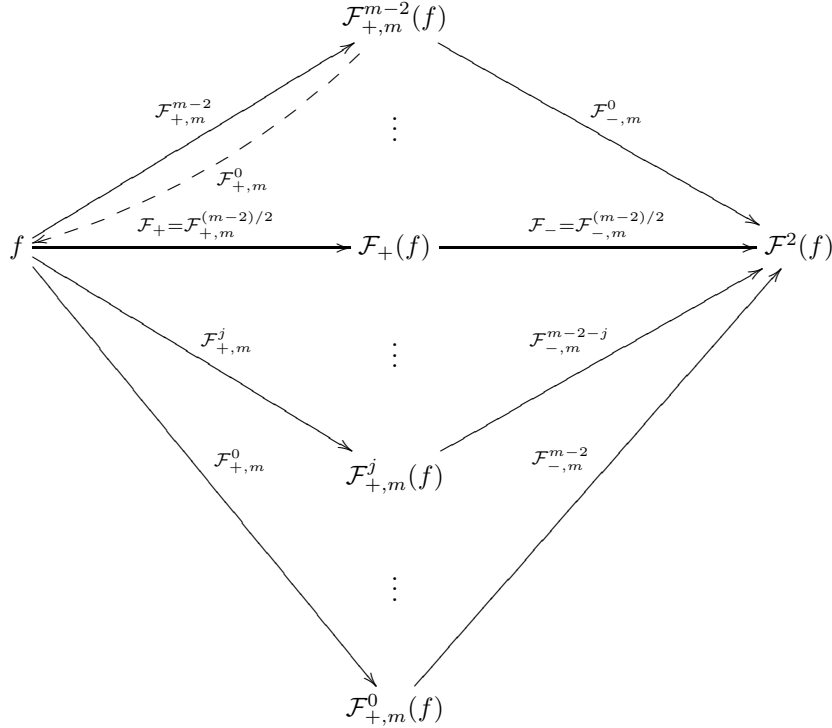
\[
\xymatrix{
&&\cF_{+,m}^{m-2}(f)\ar[rrdd]^{\cF_{-,m}^{0}} \ar@{-->}@/^1pc/[ddll]^{\cF_{+,m}^{0}}&&\\
&&\vdots&&\\
f \ar[rrdddd]^{\cF_{+,m}^{0}} \ar[rr]^{\cF_+ =\cF_{+,m}^{(m-2)/2}} \ar[rruu]^{\cF_{+,m}^{m-2}} \ar[rrdd]^{\cF_{+,m}^{j}}&\qquad \qquad \qquad&\cF_+(f) \ar[rr]^{\cF_- =\cF_{-,m}^{(m-2)/2}}&\qquad \qquad \qquad&\cF^2(f)\\
&&\vdots&&\\
&&\cF_{+,m}^{j}(f)\ar[rruu]^{\cF_{-,m}^{m-2-j}}&&\\
&&\vdots&&\\
&&\cF_{+,m}^{0}(f)\ar[rruuuu]^{\cF_{-,m}^{m-2}}&&
}
\]

The dashed line again indicates that the inverse of the transform $\mathcal{F}^{j}_{+,m}$ is given by $\mathcal{F}^{m-2-j}_{+,m}$, see formula (\ref{invCFTclass}). We may also observe that the factorization of $\cF^2$ by the CFT is not unique. In other words, only the unitarity distinguishes the CFT from the other members of this class of transforms.

\begin{remark}
Similar results clearly hold for $\widehat{K}_{\alpha,\beta}(x,y)$, the kernel of the fractional CFT
as given in Theorem \ref{EvenExplicitFr}. In this case, one has to solve the system of PDEs given in Proposition \ref{simpleSystem}. It would be worthwhile to pursue this analogy further.

\end{remark}

\subsection{Hypercomplex transforms and radial deformation}
\label{HRFTsection}

We start this section by rewriting a reflection acting on $\mR^m$ (as defined in Section \ref{DTF}) using elements from the Clifford algebra. Indeed, if for elements of our root system we identify $\alpha$ with a $1$-vector $\uA$ in $\cC l_{0, m}$ (and hence $\alpha/\sqrt{2}$ with an element in $Pin(m)$), we have
\[
r_{\alpha}(x) =\frac{1}{2} \, \uA \, \ux\, \uA
\]
with $\ux =  \sum_{i=1}^{m}e_{i}x_{i}$. Generalizing this map gives us the covering map $p$ from $Pin(m)$ to $O(m)$ as
\[
p(s)(x) = \epsilon(s) \, \ux \, s^{-1}, \quad s \in Pin(m).
\]
In particular, we obtain a double cover of the reflection group $\cG$ as $\widetilde{\cG} = p^{-1}(\cG)$ (see also the discussion in \cite{BCT}).

The starting point in the subsequent analysis relies on the Dunkl-Dirac operator, given by $\cD_{\k} = \sum_{i=1}^{m}e_{i}T_{i}$. This operator was first defined in \cite{MR2230262}.
Together with the vector variable $\ux$ this Dunkl-Dirac operator generates a copy of $\mathfrak{osp}(1|2)$, see \cite{Orsted} or the subsequent Theorem \ref{ospFamilyRad} (for $c=0$). In particular, we have
\[
\cD_{\k}^{2} = - \Delta_{\k} \quad \mbox{and} \quad \ux^{2} = - |x|^{2}=-r^{2}.
\]

\subsubsection{A new osp realization}
\label{osprel}

In this section, we will discuss the theory of radial deformations of $\mathfrak{osp}(1|2)$, as introduced in \cite{H12, DBOSS}. This theory presents a first order alternative to the radial deformation of $\mathfrak{sl}_2$ developed in \cite{Orsted2} and discussed in Section \ref{radFT}.

In the paper \cite{H12} the following family of generalized, radially deformed, Dunkl-Dirac operators was proposed:
\[
\dD = r^{1- \frac{a}{2}}\cD_{\k} + b r^{-\frac{a}{2}-1} \ux + c r^{-\frac{a}{2}-1}\ux \mE.
\]
with $a, b$ and $c$ real parameters. Subsequently, in \cite{DBOSS} it was observed that it is sufficient to study the function theory for the operator
\[
\dD = \cD_{\k}  + c r^{-2}\ux \mE,
\]
where we have put $a=2$, $b=0$, because there exist intertwining operators connecting this operator with the one depending on the three parameters $a, b$ and $c$.

Furthermore, we will restrict ourselves to the case $c> -1$ for reasons that will become clear in Proposition \ref{partIntProp}. Let us first state the fact that we indeed find a new family of realizations of $\mathfrak{osp}(1|2)$, for each value of the parameter $c$.

\begin{theorem}
\label{ospFamilyRad}
The operators $\dD$ and $\ux$ generate a Lie superalgebra, isomorphic to $\mathfrak{osp}(1|2)$, with the following relations
\begin{equation}
\begin{array}{lll}
\{ \ux, \dD\} = -2(1+c) \left( \mE + \frac{\delta}{2}\right)&\quad&\left[\mE + \frac{\delta}{2}, \dD \right] = -  \dD\\
\vspace{-3mm}\\
\left[ \ux^{2}, \dD \right] = 2(1+c)\ux&\quad&\left[\mE + \frac{\delta}{2}, \ux \right] =   \ux\\
\vspace{-3mm}\\
\left[ \dD^{2}, \ux \right] = -2(1+c)\dD&\quad&\left[\mE + \frac{\delta}{2}, \dD^{2} \right] = - 2 \dD^{2}\\
\vspace{-3mm}\\
\left[ \dD^{2}, \ux^{2} \right] = 4(1+c)^{2}\left( \mE + \frac{\delta}{2} \right)&\quad&\left[\mE + \frac{\delta}{2}, \ux^{2} \right] = 2 \ux^{2},
\end{array}
\end{equation}
where $\delta = 1 + \frac{ \mu -1}{1+c}$.
\end{theorem}

Note that the square of $\dD$ is a complicated operator, given by
\[
\dD^{2} =- \Delta_{\k} - \left( c\mu \right) r^{-1}\partial_{r}
 -  \left(c^{2} + 2c\right) \partial_{r}^{2}+c r^{-2} \sum_{i} x_{i}T_{i} -c r^{-2}  \sum_{i<j} e_{i}e_{j} ( x_{i}T_{j} - x_{j}T_{i}).
\]
If $\k =0$, the formula for $\dD^{2}$ simplifies a bit as now $ \sum_{i} x_{i}T_{i} = r \partial_{r} = \mE$. Note that the square of $\dD$ is not a scalar operator. This is very interesting from the point of view of Clifford analysis, where the square of a Dirac type operator is usually scalar.

\begin{remark}
The operator $\dD = \cD_{\k} + c r^{-2}\ux \mE$ is also considered from a very different perspective in \cite{MR2269930} for $\k =0$. In that paper, the eigenfunctions of this operator are studied.
\end{remark}

Let us now discuss the symmetry of the generators of $\mathfrak{osp}(1|2)$. First we define the action of the Pin group on $C^{\infty}(\mR^{m}) \otimes \cC l_{0, m}$ for $s \in Pin(m)$ as
\begin{align*}
\rho(s): \;\,&C^{\infty}(\mR^{m}) \otimes \cC l_{0, m} \mapsto C^{\infty}(\mR^{m}) \otimes \cC l_{0, m}\\
& f\otimes b \mapsto f (p(s^{-1})x) \otimes s b.
\end{align*}
We then have
\begin{proposition}
\label{equivarianceG}
Let $s \in \widetilde{\cG}$ and define $sgn(s):=sgn(p(s))$. Then one has
\begin{align*}
\rho(s) \, \ux &= sgn(s) \,\ux \, \rho(s)\\
\rho(s) \, \dD &= sgn(s) \,\dD \, \rho(s).
\end{align*}
\end{proposition}

So up to sign, the Dirac operator $\dD$ is $\widetilde{\cG}$-equivariant. This is the same symmetry as obtained for the Dirac operator defined in the Hecke algebra (see \cite{BCT}, Lemma 3.4).

\subsubsection{Representation space for the deformation family of the Dunkl-Dirac operator}

There is a measure naturally associated with $\dD$ given by $h(r)= r^{1- \frac{1+\mu c}{1+c}}$. One finds, after tedious computations in \cite{H12},

\begin{proposition}
\label{partIntProp}
If $c > -1$, then for suitable differentiable functions $f$ and $g$ one has
\[
\int_{\mR^{m}} \overline{(\dD f)} \; g \; h(r) w_{\kappa}(x) dx = \int_{\mR^{m}} \overline{f} \; (\dD g) \; h(r) w_{\k}(x) dx
\]
with $h(r) =  r^{1- \frac{1+\mu c}{1+c}}$, provided the integrals exist.
\end{proposition}
In this proposition, $\bar{.}$ is the main anti-involution on the Clifford algebra $\cC l_{0, m}$.

The function space we will use to further develop the theory is $\cL^{2}_{\k,c}(\mR^{m}) := L^{2}(\mR^{m}, h(r) w_{\k}(x)dx) \otimes \cC l_{0, m}$. This space has the following decomposition
\[
\cL^{2}_{\k,c}(\mR^{m})  =  L^{2}(\mR^{+}, r^{\frac{\mu-1}{1+c}}dr) \otimes L^{2}(\mS^{m-1},w_{\k}(\xi) d\sigma(\xi)) \otimes \cC l_{0, m}
\]
where on the right-hand side the topological completion of the tensor product is understood and with $d\sigma(\xi)$ the Lebesgue measure on the sphere $\mS^{m-1}$. The space $ L^{2}(\mS^{m-1},w_{\k}(\xi) d\sigma(\xi)) \otimes \cC l_{0, m}$ can be further decomposed into Dunkl harmonics and subsequently into Dunkl monogenics.
This leads to
\[
L^{2}(\mS^{m-1},w_{\k}(\xi) d\sigma(\xi)) \otimes \cC l_{0, m} = \bigoplus_{\ell=0}^{\infty} \left. \left(  \cM_{\ell}^{\cD} \oplus \ux   \cM_{\ell}^{\cD} \right) \right|_{\mS^{m-1}},
\]
where $\cM_{\ell}^{\cD} := \ker{\cD_{\k}} \cap  \left(\cP_{\ell} \otimes \cC l_{0, m} \right)$ is the space of Dunkl monogenics of degree $\ell$. For more details on Dunkl monogenics, we refer the reader to \cite{DBGeg}.

 Using this decomposition, a basis for $\cL^{2}_{\k,c}(\mR^{m})$ was obtained in \cite{H12}. This basis is given by the set $\{ \psi^c_{t,\ell,m}\}$ ($t, \ell \in \mN$ and $m=1, \ldots, \dim \cM_{\ell}^{\cD} $), defined as
\begin{align*}
\psi^c_{2t,\ell,m} &:= 2^{2t} (1+c)^{2t} t! L_{t}^{\frac{\gamma_{\ell}}{2} -1}(r^{2})r^{\beta_{\ell}} M_{\ell}^{(m)} e^{-r^{2}/2},\\
\psi^c_{2t+1,\ell,m} &:=  - 2^{2t+1} (1+c)^{2t+1} t!  L_{t}^{\frac{\gamma_{\ell}}{2}}(r^{2}) \ux r^{\beta_{\ell}} M_{\ell}^{(m)} e^{-r^{2}/2}
\end{align*}
with $L_{\alpha}^{\beta}$ the Laguerre polynomials and
\begin{align*}
\beta_{\ell} &= - \frac{c}{1+c}\ell ,\\
\gamma_{\ell} &= \frac{2}{1+c}\left( \ell + \frac{\mu-2}{2}\right) + \frac{c+2}{1+c}
\end{align*}
and where $M_{\ell}^{(m)}$ ($m=1, \ldots, \dim \cM_{\ell}^{\cD}$) forms an orthonormal basis of $\cM_{\ell}^{\cD}$, i.e. 
\[
\left[ \int_{\mS^{m-1}}\overline{M_{\ell}^{(m_{1})}}(\xi) M_{\ell}^{(m_{2})}(\xi) w_{\k}(\xi) d\sigma(\xi) \right]_{0} = \delta_{m_{1} m_{2}}.
\] 
The dimension of $\cM_{\ell}^{\cD}$ is given by
\begin{align*}
\dim_{\mR}{\cM_{\ell}} &= \dim_{\mR}{\cC l_{0, m}} \dim_{\mR}{\cP_{\ell}(\mR^{m-1})}\\
&=2^{m} \frac{(\ell +m-2)!}{\ell! (m-2)!}
\end{align*}
with $\cP_{\ell}(\mR^{m-1})$ the space of homogeneous polynomials of degree $\ell$ in $m-1$ variables (see \cite{MR1169463}). Note that for $c=0$ and $\k=0$, the basis $\{ \psi^c_{t,\ell,m} \}$ reduces to the basis $\{ \psi_{t,\ell,m} \}$ introduced in Section \ref{prelims}.

Using formula (4.10) in \cite{H12} and the proof of Theorem 3 in \cite{H12}, one obtains the following formulas for the action of $\dD$ and $\ux$ on the generalized Laguerre functions
\begin{align}
\label{ActionOnBasis}
\begin{split}
2 \dD \psi^c_{t,\ell,m} & = \psi^c_{t+1,\ell,m} + C(t,\ell) \psi^c_{t-1,\ell,m},\\
-2(1+c) \ux \psi^c_{t,\ell,m} & = \psi^c_{t+1,\ell,m} - C(t,\ell) \psi^c_{t-1,\ell,m}
\end{split}
\end{align}
with
\begin{align*}
C(2t, \ell)&= 4 (1+c)^{2} t,\\
C(2t+1, \ell)&= 2 (1+c)^{2}(\gamma_{\ell}+ 2t).
\end{align*}
These formulas determine the action of $\mathfrak{osp}(1|2)$ on $\cL^{2}_{\k,c}(\mR^{m})$. Recall also that the action of $\widetilde{\cG}$ on $\cL^{2}_{\k,c}(\mR^{m})$ is given by $\rho$ (see Section \ref{osprel}). 

The action of $\mathfrak{osp}(1|2)$ can be made even more explicit by defining creation and annihilation operators via
\begin{align}
\label{CreaAnn}
\begin{split}
A^{+} &= \dD - (1+c) \ux ,\\
A^{-} &= \dD + (1+c) \ux.
\end{split}
\end{align}
These operators satisfy 
\begin{align*}
A^{+} \psi^c_{t,\ell,m}& = \psi^c_{t+1,\ell,m} ,\\
A^{-} \psi^c_{t,\ell,m}& = C(t, \ell) \psi^c_{t-1,\ell,m}.
\end{align*}

Now we introduce the following inner product
\[
\langle f, g \rangle = \left[ \int_{\mR^{m}} \overline{f^{c}} \, g \;h(r) w_{\k}(x) dx \right]_{0}
\]
where $h(r)$ is the measure associated to $\dD$ (see Proposition \ref{partIntProp}) and $f^{c}$ is the complex conjugate of $f$. It is easy to check that this inner product satisfies
\begin{align*}
\langle \dD f, g \rangle &= \la f, \dD g\ra\\
\la \ux f, g \ra &= - \la f, \ux g \ra.
\end{align*}
The related norm is defined by $||f||^{2} = \la f,f \ra$. Using this inner product, it was possible to prove the orthogonality of the new basis $\{ \psi^c_{t,\ell,m} \}$.

\begin{theorem}
\label{orthLaguerre}
One has
\[
\langle \psi^c_{t_{1},\ell_{1}, m_{1}} , \psi^c_{t_{2},\ell_{2}, m_{2}} \rangle = c(t_{1},\ell_{1}) \delta_{t_{1} t_{2}} \delta_{\ell_{1} \ell_{2}}\delta_{m_{1} m_{2}} 
\]
where $c(t, \ell)$ is a constant depending on $t$ and $\ell$.
\end{theorem}
Explicit expressions for the constants $c(t, \ell)$ have been obtained in \cite{H12}.

The functions $\psi^c_{t,\ell,m}$ are eigenfunctions of the hamiltonian of a generalized harmonic oscillator.
\begin{theorem}
\label{HarmOsc}
The functions $\psi^c_{t,\ell,m}$ satisfy the following second-order PDE
\[
\left(\dD^{2} - (1+c)^{2}\ux^{2}\right) \psi^c_{t,\ell,m} = (1+c)^{2}(\gamma_{\ell} + 2 t ) \psi^c_{t,\ell,m}.
\]
\end{theorem}

Theorem \ref{HarmOsc} combined with the definition of $A^{+},A^{-}$ in (\ref{CreaAnn}) allows us to decompose the space $\cL^{2}_{\k,c}(\mR^{m})$ under the action of $\mathfrak{osp}(1|2)$. Clearly the odd elements $A^{+}$ and $A^{-}$ generate  
$\mathfrak{osp}(1|2)$ as they are linear combinations of $\dD$ and $\ux$. Moreover, they act between basis vectors 
of $\cL^{2}_{\k,c}(\mR^{m})$, so it is sufficient to consider such vectors in an irreducible
representation of $\mathfrak{osp}(1|2)$ inside the function space. This is achieved as follows: for fixed $\ell$ and $m$ each vector $\psi^c_{0,\ell,m}$ generates the irreducible representation
\[
\xymatrix{
\psi^c_{0,\ell,m} \ar@<1ex>[r]^{A^+} \ar@(dl,dr)_{L}   & \psi^c_{1,\ell,m} \ar@(dl,dr)_{L} \ar@<1ex>[r]^{A^+} \ar@<1ex>[l]^{A^-}   & \psi^c_{2,\ell,m} \ar@(dl,dr)_{L} \ar@<1ex>[r]^{A^+} \ar@<1ex>[l]^{A^-} &\psi^c_{3,\ell,m} \ar@(dl,dr)_{L} \ar@<1ex>[r]^{A^+} \ar@<1ex>[l]^{A^-} &\psi^c_{4,\ell,m}  \ar@(dl,dr)_{L} \ar@<1ex>[r]^{A^+} \ar@<1ex>[l]^{A^-} &\ar@<1ex>[l]^{A^-} \ldots
}
\]
where
\[
L = \frac{1}{2}\{A^{+},A^{-}\} = \dD^{2} - (1+c)^{2} \ux^{2}
\]
with the action given in Theorem \ref{HarmOsc}.
In fact this highest weight representation is labeled by $\ell$ only and we can denote it $\pi(\ell)$.  
In conclusion, we obtain the decomposition of the space $\cL^{2}_{\k,c}(\mR^{m})$ 
into a discrete direct sum of highest weight (infinite dimensional)
Harish-Chandra modules for $\mathfrak{osp}(1|2)$: 
\[
\cL^{2}_{\k,c}(\mR^{m}) = \bigoplus_{\ell=0}^{\infty} \pi(\ell) \otimes \cM_{\ell}^{\cD}.
\]

These results should be compared with Theorem 3.19 and section 3.6 in \cite{Orsted2} (where one uses $\mathfrak{sl}_{2}$ instead of $\mathfrak{osp}(1|2)$). Also notice that the claim should be understood as an assertion on the deformation of the Howe dual pair 
for $\mathfrak{osp}(1|2)$ inside the Clifford-Weyl algebra on $\mR^{m}$ acting on a fixed vector space $\cL^{2}_{\k,c}(\mR^{m})$.

\begin{remark}
One can also consider more general deformations of the Dirac operator, by adding suitable odd powers of $\Gamma_x = -\ux \cD_{\k}-\mE$ to $\dD$ as follows
\[
\dD = \cD_{\k}  + c r^{-2}\ux \mE + \sum_{j=0}^{\ell} c_{j} r^{-1}\left(\Gamma_x-\frac{\mu-1}{2} \right)^{2j+1}, \qquad c_{j} \in \mR.
\]
This does not alter the $\mathfrak{osp}(1|2)$ relations, as $\Gamma_x -\frac{\mu-1}{2} $ anti-commutes with $\ux$ and has the correct homogeneity. In particular, $\Gamma_x -\frac{\mu-1}{2} $ can be seen as the square root of the Casimir of $\mathfrak{osp}(1|2)$, see \cite{MR1773773}, example 2 in section 2.5.
\end{remark}

In the rest of Section \ref{HRFTsection}, we will always assume $\kappa=0$ or in other words, we do not consider the Dunkl deformation. This is to simplify the notation of the results. Most statements can immediately be generalized to the Dunkl case by composition with the intertwining operator $V_{\k}$. Recall that for $\kappa=0$, the Dunkl-Dirac operator $\cD_{\k}$ reduces to the orthogonal Dirac operator $\upx=\sum_{i=1}^{m} e_{i} \partial_{x_{i}}$.

\subsubsection{Clifford deformed Hermite semigroup}

The aim of the present section is to discuss the properties of the holomorphic semigroup defined by
\[
\cF_{\dD}^{\o} = e^{\o \left(\frac{1}{2}+ \frac{\mu - 1}{2(1+c)} \right)} e^{\frac{-\o}{ 2(1+c)^{2}}\left(\dD^{2} - (1+c)^{2}\ux^{2}\right)}, \quad \Re \o \geq 0
\]
acting on the space $\cL^{2}_{0,c}(\mR^{m})$. We start with the following general statement, which is obtained by considering the basis $\{ \psi^c_{t,\ell,m} \}$.

\begin{theorem}
\label{PropSemigroup}
Suppose $c > -1$. Then
\begin{enumerate}
\item
For any $t, \ell \in\mN$ and $m \in \{1, \ldots, \dim \cM_{\ell}\}$, the function $\psi^c_{t,\ell,m}$ is an eigenfunction of the 
operator $\cF_{\dD}^{\o}$: 
$$
\cF_{\dD}^{\o} (\psi^c_{t,\ell,m}) = e^{-\o t} e^{-\frac{\o \ell}{(1+c)}}\psi^c_{t,\ell,m}.
$$
\item $\cF_{\dD}^{\o}$ is a continuous operator on $\cL^{2}_{0,c}(\mR^{m})$ for all $\o$ with $\Re \o \geq 0$, as $||\cF_{\dD}^{\o}(f)|| \leq ||f||$
for all $f \in \cL^{2}_{0,c}(\mR^{m})$.
\item
If $\Re \o > 0$, then $\cF_{\dD}^{\o}$ is a Hilbert-Schmidt operator on $\cL^{2}_{0,c}(\mR^{m})$.
\item
If $\Re \o = 0$, then $\cF_{\dD}^{\o}$ is a unitary operator on $\cL^{2}_{0,c}(\mR^{m})$.
\end{enumerate} 
\end{theorem}

Let us now derive a series expansion for the integral kernel in the sense of \textbf{F4}. We summarize the result in the following theorem. Here we use again the notation $\widetilde{J}_{\nu}(z) = (z/2)^{-\nu}J_{\nu}(z)$ and $r^{2}= |x|^{2}$,  $s^{2}= |y|^{2}$, as well as $\l = (m-2)/2$.

\begin{theorem}
\label{seriesholom}
Let $\Re \o > 0$ and $c > -1$.
Put $K(x,y;\omega) =  e^{-\frac{\coth{\o}}{2} (r^{2}+ s^{2})} \left(A(z,w) + \ux \wedge \uy B(z,w)\right)
$ with
\begin{align*}
A(z,w) &= \sum_{k=0}^{+\infty}  \left(\alpha_{k} \frac{k+2\l}{2\l} z^{\frac{k}{1+c}} \widetilde{J}_{\frac{\gamma_{k}}{2}-1}\left(\frac{iz}{\sinh{\o}}\right)  +\frac{ \alpha_{k-1}}{4\sinh{\o}} \frac{k}{\l} z^{\frac{k+c}{1+c}} \widetilde{J}_{\frac{\gamma_{k-1}}{2}}\left(\frac{iz}{\sinh{\o}}\right)\right)  C_{k}^{\l}(w),\\
B(z,w) &= \sum_{k=1}^{+\infty} \left(-\alpha_{k} z^{\frac{k}{1+c}-1}  \widetilde{J}_{\frac{\gamma_{k}}{2}-1}\left(\frac{iz}{\sinh{\o}}\right)  + \frac{\alpha_{k-1}}{2\sinh{\o}} z^{\frac{k+c}{1+c}-1} \widetilde{J}_{\frac{\gamma_{k-1}}{2}}\left(\frac{iz}{\sinh{\o}}\right)\right) C_{k-1}^{\l+1}(w)
\end{align*}
for $z = |x| |y|$, $w = \la x, y \ra /z$, $\alpha_{-1}=0$ and $\alpha_{k} = 2 e^{\frac{\o \d}{2}}  (2 \sinh{\o})^{-\gamma_{k}/2}$.
Then these series are convergent and the integral transform defined on $\cL^{2}_{0,c}(\mR^{m})$ by
\[
\cF_{0, c}^{\o}(f)(y) = \sigma_{m}^{-1}\int_{\mR^{m}} K(x,y;\omega) f(x) h(r) dx
\]
coincides with the operator $\cF_{\dD}^{\o} = e^{\o \left(\frac{1}{2}+ \frac{\mu - 1}{2(1+c)} \right)} e^{\frac{-\o}{ 2(1+c)^{2}}\left(\dD^{2} - (1+c)^{2}\ux^{2}\right)}$ on the basis $\{ \psi^c_{t,\ell,m} \}$.
\end{theorem}

We are also interested in the case where the semigroup parameter satisfies $\Re \o = 0$. In this case, one takes the limit to the imaginary axis and obtains the following result.

\begin{theorem}
\label{seriesunitary}
Let $c > -1$. Then for $\o = i \eta$ with $\eta \not \in \pi \mZ$, we put $K(x,y;i \eta) =  e^{i\frac{\cot{\eta}}{2} (r^{2}+ s^{2})} \left(A(z,w) + \ux \wedge \uy B(z,w)\right)$
with
\begin{align*}
A(z,w) &= \sum_{k=0}^{+\infty}  \left(\alpha_{k} \frac{k+2\l}{2\l} z^{\frac{k}{1+c}} \widetilde{J}_{\frac{\gamma_{k}}{2}-1}\left(\frac{z}{\sin{\eta}}\right)  +\frac{ \alpha_{k-1}}{4i\sin{\eta}} \frac{k}{\l} z^{\frac{k+c}{1+c}} \widetilde{J}_{\frac{\gamma_{k-1}}{2}}\left(\frac{z}{\sin{\eta}}\right)\right)  C_{k}^{\l}(w),\\
B(z,w) &= \sum_{k=1}^{+\infty} \left(-\alpha_{k} z^{\frac{k}{1+c}-1}  \widetilde{J}_{\frac{\gamma_{k}}{2}-1}\left(\frac{z}{\sin{\eta}}\right)  + \frac{\alpha_{k-1}}{2i\sin{\eta}} z^{\frac{k+c}{1+c}-1} \widetilde{J}_{\frac{\gamma_{k-1}}{2}}\left(\frac{z}{\sin{\eta}}\right)\right) C_{k-1}^{\l+1}(w)
\end{align*}
for $z = |x| |y|$, $w = \la x, y \ra /z$, $\alpha_{-1}=0$ and $\alpha_{k} = 2 e^{\frac{i \eta \d}{2}}  (2 i\sin{\eta})^{-\gamma_{k}/2}$.
These series are convergent and the unitary integral transform defined in distributional sense on $\cL^{2}_{0,c}(\mR^{m})$ by
\[
\cF_{0, c}^{i\eta}(f)(y) = \sigma_{m}^{-1}\int_{\mR^{m}} K(x,y;i\eta) f(x) h(r) dx
\]
coincides with the operator $\cF_{\dD}^{i\eta} = e^{i\eta \left(\frac{1}{2}+ \frac{\mu - 1}{2(1+c)} \right)} e^{\frac{-i\eta}{ 2(1+c)^{2}}\left(\dD^{2} - (1+c)^{2}\ux^{2}\right)}$ on the basis $\{ \psi^c_{t,\ell,m} \}$.
\end{theorem}

\subsubsection{The radially deformed hypercomplex Fourier transform}

The Fourier transform is the very special case of the holomorphic semigroup, evaluated at $\o = i \pi /2$. In this case, the kernel $K(x,y)= K(x,y;i\pi/2)$ is given by the following theorem. In the limit case $c=0$, we can check that the result reduces to expression \textbf{F4}, as expected.

\begin{theorem}
\label{seriesrepr}
Put $K(x,y) = A(z,w) + \ux \wedge \uy B(z,w)$ with
\begin{align*}
A(z,w) &= \sum_{k=0}^{+\infty}  z^{-\frac{\delta-2}{2}} \left(\alpha_{k} \frac{k+2\l}{2\l} J_{\frac{\gamma_{k}}{2}-1}(z)  -i \alpha_{k-1} \frac{k}{2\l} J_{\frac{\gamma_{k-1}}{2}}(z)\right)  C_{k}^{\l}(w),\\
B(z,w) &= \sum_{k=1}^{+\infty} z^{-\frac{\delta}{2}} \left(-\alpha_{k} J_{\frac{\gamma_{k}}{2}-1}(z)  -i \alpha_{k-1} J_{\frac{\gamma_{k-1}}{2}}(z)\right) C_{k-1}^{\l+1}(w)
\end{align*}
and $z = |x| |y|$, $w = \la x, y \ra /z$, $\alpha_{-1}=0$ and $\alpha_{k} = e^{-\frac{i \pi k}{2(1+c)}}$.
These series are convergent and the integral transform defined in distributional sense on $\cL^{2}_{0,c}(\mR^{m})$ by
\[
\cF_{0, c}(f)(y) = \sigma_{m}^{-1}\int_{\mR^{m}} K(x,y) f(x) h(r) dx
\]
coincides with the operator $\cF_{\dD} = e^{i \frac{\pi}{2} \left(\frac{1}{2}+ \frac{\mu - 1}{2(1+c)} \right)} e^{\frac{-i \pi}{ 4(1+c)^{2}}\left(\dD^{2} - (1+c)^{2}\ux^{2}\right)}$
on the basis $\{ \psi^c_{t,\ell,m} \}$.
\end{theorem}

\begin{remark}
One can also define an analogue of the Schwartz space of rapidly decreasing functions in this context. Let $L = \dD^{2} - (1+c)^{2}\ux^{2}$ and denote by $D(L)$ the domain of $L$ in  $\cL^{2}_{0,c}(\mR^{m})$. Then the Schwartz space is defined by
\[
\cS_{0,c}(\mR^{m}) = \bigcap_{k=0}^{\infty} D(L^{k})
\]
and one can check that the Fourier transform $\cF_{0, c}$ is an isomorphism of this space.
\end{remark}

The symmetry properties of the integral kernel are discussed in the following proposition.
\begin{proposition}
One has, with $x,y \in \mR^{m}$
\begin{align*}
K(\l x,y) &= K(x, \l y), \quad \l > 0\\
K(y,x) &= \overline{K(x,y)},\\
K(0,y) & = \frac{1}{2^{\gamma_{0}/2-1}\Gamma(\gamma_{0}/2)},\\
K(\overline{s} x s,\overline{s} y s) &= \overline{s} K(x,y) s, \quad s \in Spin(m).
\end{align*}
where $\bar{.}$ is the anti-involution on the Clifford algebra $\cC l_{0, m}$.
\end{proposition}

We can now summarize the main properties of the deformed Fourier transform in the following theorem.

\begin{theorem}
\label{StatementsFourier}
The operator $\cF_{0, c}$ defines a unitary operator on $\cL^{2}_{0,c}(\mR^{m})$ and satisfies the following intertwining relations on a dense subset:
\begin{align*}
\cF_{0, c} \circ \dD &=  i (1+c) \ux \circ \cF_{0, c},\\
\cF_{0, c} \circ \ux &= \frac{i}{1+c}\dD \circ \cF_{0, c},\\
\cF_{0, c} \circ \mE &= - \left( \mE + \delta \right) \circ \cF_{0, c}.
\end{align*}
Moreover, $\cF_{0, c}$ is of finite order if and only if $1+c$ is rational.
\end{theorem}

We can also obtain Bochner identities for the deformed Fourier transform. They are given in the following proposition.

\begin{proposition}
\label{Bochner}
Let $M_{\ell} \in \cM_{\ell}$ be a spherical monogenic of degree $\ell$. Let $f(x)= f(r)$ be a radial function. Then the Fourier transform of $f(r) M_{\ell}$ and $f(r) \ux M_{\ell}$ can be computed as follows:
\begin{align*}
\cF_{0, c}\left(f(r) M_{\ell}\right) &= e^{-\frac{i \pi \ell}{2(1+c)}} M_{{\ell}}(\uy') \int_{0}^{+\infty} r^{\ell} f(r) z^{-\frac{\delta-2}{2}} J_{\frac{\gamma_{k}}{2} -1}(z)  h(r) r^{m-1} dr ,\\
\cF_{0, c} \left(f(r) \ux M_{\ell}\right) &= -i e^{-\frac{i \pi \ell}{2(1+c)}} \uy' M_{{\ell}}(\uy') \int_{0}^{+\infty} r^{\ell+1} f(r) z^{-\frac{\delta-2}{2}} J_{\frac{\gamma_{k}}{2} }(z)  h(r)r^{m-1} dr
\end{align*}
with $y = s y'$, $y' \in \mS^{m-1}$ and $z = r s$. 
\end{proposition}

The Heisenberg inequality (or uncertainty principle) is a very important qualitative statement about the ordinary Fourier transform, see e.g. \cite{Fol} for a review. There exist extensions to the Dunkl transform \cite{MR1698045, MR1818904}, the radially deformed Fourier transform \cite{Orsted2} and the super Fourier transform \cite{CDB}. Also in the present context, it is possible to obtain a Heisenberg inequality. This is the subject of the following proposition, obtained in \cite{DBOSS}.

\begin{proposition}
\label{Heisenberg}
For all $f \in \cL^{2}_{0,c}(\mR^{m})$, the radially deformed hypercomplex Fourier transform satisfies
\[
|| \ux \, f(x)|| . || \ux \, \cF_{0, c}(f)(x)|| \geq \frac{\d}{2} ||f(x)||^{2},
\]
with $\delta$ defined in Theorem \ref{ospFamilyRad}.
The equality holds if and only if $f$ is of the form $f(x) = \l e^{-r^{2}/\a}$.
\end{proposition}

For the CFT, discussed in section \ref{CFTsection}, currently no Heisenberg inequality is available. Nevertheless, its proof should proceed along similar lines.
Note that also for other hypercomplex and quaternionic Fourier transforms, as discussed in the subsequent Section \ref{OtherTFsSect}, uncertainty principles have been obtained \cite{Maw1, Maw2, Maw4, Maw3}. These results have however been proved using a completely different strategy.

Let us end this section with an important result, the so-called Cherednik master formula for the kernel of the semigroup. This forms the starting point for the study of a generalized heat equation, see e.g. \cite[Lemma 4.5 (1)]{MR1620515} in the context of Dunkl operators. We may state the master formula as follows:

\begin{theorem}[Master formula]
\label{MasterFormula}
Let $s > 0$. Then one has
\[
\int_{\mR^{m}} K(y,x;i\frac{\pi}{2}) K(z,y; - i\frac{\pi}{2}) e^{-s |y|^{2}} h(|y|) dy = \s_{m} e^{-\frac{\o \d}{2}} K(z,x; \o) e^{-\frac{|x|^{2}+|z|^{2}}{2}\frac{1- \cosh{\o}}{\sinh{\o}}}
\]
with $2 s = \sinh{\o}$.
\end{theorem}

In the proof of this theorem, one uses the integral formula (see \cite[p. 50]{Erde})
\begin{equation}
\label{besselint1}
\int_{0}^{+\infty} J_{\nu}(a t)J_{\nu}(b t) e^{-\gamma^{2}t^{2}} t dt = \frac{1}{2} \gamma^{-2} e^{-\frac{a^{2}+b^{2}}{4 \gamma^{2}}} I_{\nu}( \frac{a b}{2 \gamma^{2}}), \quad \Re \nu > -1, \Re \gamma^{2}> 0
\end{equation}
where $I_{\nu}(z) = e^{-i \frac{\pi \nu}{2}} J_{\nu}(i z)$. For the Dunkl transform (see e.g. \cite{MR1973996, MR2274972}) and for the Clifford-Fourier transform (see \cite{DBXu}) one can even compute a more general integral of the form
\[
\int_{\mR^{m}} K(y,x;i\frac{\pi}{2}) K(z,y; - i\frac{\pi}{2}) f(|y|) h(|y|) dy
\]
with $f(|y|)$ an arbitrary radial function of suitable decay. This is achieved by using the addition formula for the Bessel function 
\[
u^{-\l} J_{\l}(u)= 2^{\l} \Gamma(\l)\sum_{k=0}^{\infty}  (k+\l)(r^{2} |x||z|)^{-\l} J_{k+\l}(r |x|)J_{k+\l}(r |z|) C^{\l}_{k}(\la x' , z' \ra)
\]
with $u = r \sqrt{|x|^{2}+|z|^{2} - 2\la x,z\ra}$ instead of formula (\ref{besselint1}). Here, we cannot do that, as the orders of the Bessel functions do not match the orders of the Gegenbauer polynomials in a suitable way.

\subsubsection{Conceptual summary of deformations}

It now becomes possible to depict the various deformations of the Fourier transform, both scalar and hypercomplex, in a diagram.  In this diagram, the hamiltonian of the harmonic oscillator is first deformed using the Dunkl Laplacian (giving rise to a CMS quantum system). At this point, two conceptual radial deformations exist. Either one deforms the Dunkl Laplacian with a radial factor, giving rise to the radially deformed Fourier transform studied in \cite{Orsted2} and Section \ref{radFT} or one deforms instead the underlying Dunkl-Dirac operator as done in \cite{H12, DBOSS} and discussed in the present section.
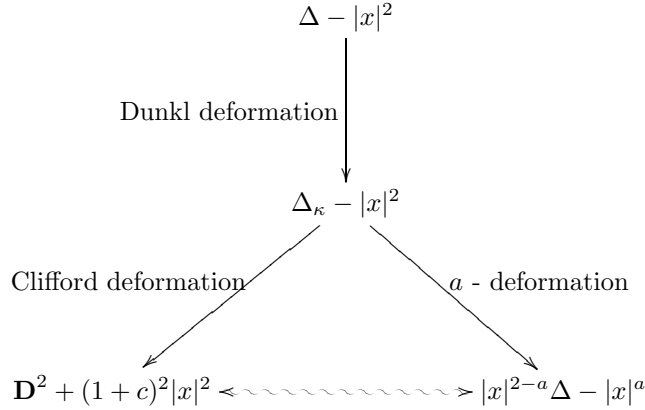
\[
\xymatrix{
&\Delta -|x|^{2}\ar[dd]_{\mbox{Dunkl deformation}}&\\
\\
&\Delta_{\k}- |x|^{2}\ar[ddl]_{\mbox{Clifford deformation}}\ar[ddr]^{\mbox{$a$ - deformation}}&\\
\\
\dD^{2}+(1+c)^{2}|x|^{2} \ar@{<~>}[rr]&&|x|^{2-a}\Delta - |x|^{a} 
}
\]

\subsection{Other approaches to hypercomplex Fourier transforms}
\label{OtherTFsSect}

Several other generalizations of the Fourier transform in the setting of quaternionic or Clifford analysis have been devised during the last 30 years. These transforms are not defined by exponentiating a differential operator or quantum hamiltonian and are hence not related to a realization of $\mathfrak{osp}(1|2)$. Instead, they are defined by replacing the complex unit $i$ in the definition of the classical Fourier transform by a root of $-1$ that belongs to a certain Clifford algebra. The aim of the present paper is not to present a detailed review of all the work that has been done in this direction, as good and detailed reviews are already available (see e.g. \cite{AIEP}). Nevertheless, we want to present the main features of these types of transforms for the sake of completeness. We start from their most general expression, which was introduced in \cite{BU}. 

Consider the general Clifford algebra $\cC l_{p,q}$ over $\mR^{p,q}$, generated by $e_{i}$, $i= 1, \ldots, p+q$, under the relations
\begin{align*} 
\begin{split}
&e_{i} e_{j} + e_{j} e_{i} = 0, \qquad i \neq j,\\
& e_{i}^{2} = 1, \quad i \leq p\\
&e_{i}^{2} =-1, \quad i > p.
\end{split}
\end{align*}
Denote further by $\cI_{p,q}$ the set $\{ a \in \cC l_{p,q} | a^{2}\in \mR^{-} \}$. A geometric Fourier transform (GFT) according to \cite{BU} is then defined by two ordered finite sets $F_{1} =\{ i_{1}(x,y), \ldots, i_{\mu}(x,y) \}$ and $F_{2} =\{ i_{\mu +1}(x,y), \ldots, i_{\mu + \nu}(x,y) \}$ of mappings $i_{k}(x,y): \mR^{m}\times \mR^{m} \rightarrow \cI_{p,q}$, by
\begin{equation}
\label{GFT}
\cF_{F1, F2} (f)(y):= \int_{\mR^{m}} \left( \prod_{j=1}^{\mu} e^{i_{j}(x,y)} \right) f(x)  \left(\prod_{j=\mu+1}^{\mu +\nu} e^{i_{j}(x,y)} \right) dx
\end{equation}
where $f: \mR^{m} \rightarrow \cC l_{p,q}$.

Without claiming completeness, we now present a few examples of GFTs that have received considerable attention in the literature, before the general definition was stated in \cite{BU}.

The first example of a GFT was given by Sommen in \cite{MR650399, MR690496} and studied further by B\"ulow and Sommer in e.g. \cite{MR1875365}. In this case, one considers the Clifford algebra $\cC l_{0,m}$. The transform takes the form
\[
\cF (f)(y):= \int_{\mR^{m}}  f(x) \,e^{e_{1} x_{1}y_{1}} \ldots  e^{e_{m} x_{m}y_{m}} dx.
\]
Derivation properties of this transform can e.g. be found in \cite{MR697564}.

The quaternionic Fourier transform was defined by Ell in \cite{EPHD, E1} and used for color image processing in e.g. \cite{MR2460142}. Considering $\mu, \nu \in \cC l_{0,2} \cong \mH$ with $\mu^2=\nu^2=-1$, it is defined via
\[
\cF^{\mu,\nu}(h) (y_1,y_2) := (2 \pi)^{-1} \int_{\mR^2} e^{-\mu x_1 y_1} \,h(x_1,x_2) \,e^{-\nu x_2 y_2} dx_1 dx_2, \quad h \in L^1(\mR^2; \mH)
\]
and is again a special case of the GFT.

Another example is the Clifford Fourier transform (written without hyphen) introduced by Ebling and Scheuermann in \cite{Ebl2} and extended by Bahri and Hitzer \cite{Maw1, Maw3}. This transform is defined for $\cC l_{m, 0}$ with $m=2 \mod{4}$ or $m=3 \mod{4}$ by
\[
\cF (f)(y):= \int_{\mR^{m}}  f(x) \,e^{e_{1 \ldots m} \langle x,y \rangle}  dx.
\]
with $e_{1 \ldots m} = e_{1} e_{2} \ldots e_{m}$ the pseudoscalar in $\cC l_{m,0}$.

Finally, the so-called cylindrical Fourier transform is defined for $\cC l_{0,m}$ by
\[
\cF (f)(y):= \int_{\mR^{m}}  e^{\, \ux \wedge \uy} \,f(x) dx.
\]
A detailed study of this transform can be found in \cite{AIEP}. Note that for dimension $m=2$ this transform coincides with the Clifford-Fourier transform discussed in Section \ref{CFTsection}.

Many properties of the GFT, defined in equation (\ref{GFT}), have been derived in general in \cite{BU}. Moreover, under some conditions on the sets $F_{1}$ and $F_{2}$ a convolution theorem was obtained in \cite{BU2}.

Finally, not all hypercomplex transforms studied so far are covered in definition (\ref{GFT}). Most notably, the transform obtained by considering the monogenic extension of the usual exponential kernel is not included. For a detailed study, we refer the reader to the papers \cite{MR2200115, MR1887633, MR1308706}.

\section{Open problems and further research}

In the final section of this paper, we indicate some open problems and present some directions for future research.

Probably the most important problem is to find explicit closed formulas for the integral kernels that are only available as a series expansion. These include the radially deformed FT, which is currently only known in closed form for $a=1$ and $a=2$. Also for the Clifford-Fourier transform, the kernel is not known explicitly in the case of odd dimension. Finally, the kernel of the radially deformed hypercomplex FT is only known explicitly for $c=0$ (which corresponds with the ordinary FT). Especially to obtain sharp bounds on the kernel functions, it is crucial to have such explicit formulas available.

It seems that finding these closed formulas is a hard problem, as there is no transparent method available for that purpose. Rather, for each kernel a combination of ad-hoc methods is necessary. One possibility would be to try and find a method that allows to compute bounds on kernel functions without needing the closed formula (in analogy with the work done for the Dunkl transform). An additional difficulty is that there seems to be a huge difference between the even and odd dimensional cases. This can e.g. be read from Lemma \ref{RecursionsProps}, where a recursion on the dimension of size 2 is observed. The odd dimensional case turns out to be much harder, because one cannot go from dimension $m=1$ (where the kernel is usually explicitly known) to dimension $m=3$. 

Also the kernel of the Dunkl transform is not known explicitly for most finite reflection groups. This is less problematic, as one has already been able to prove boundedness of the kernel using different methods. Nevertheless, it would be insightful (but very difficult) to find the Dunkl kernel in closed form for some classes of reflection groups.

Another interesting topic for further investigation is that of translation and convolution for hypercomplex Fourier transforms. For generalized Fourier transforms, there are two conceptual ways to introduce an associated convolution product. One can follow the strategy taken for the Dunkl transform (see e.g. \cite{MR2274972}), where convolution is defined using a generalized translation operator. Alternatively, one can define convolution directly as
\[
f * g := \cF^{-1}\left( \cF(f) \cF(g)\right)
\]
with $\cF$ and $\cF^{-1}$ a generalized FT and its inverse. This has already been pursued for the fractional FT, see \cite{MR1668139}. 

It seems that a detailed study of all possible translation and convolution structures for the hypercomplex FTs mentioned in this paper would be worthwhile. In particular, also a comparison with ordinary convolution (and correlation) as used e.g. in  \cite{Moxey} for the quaternionic FT would be interesting from the point of view of applications.

Next,  the function theory for the radially deformed Dirac operator should be further developed, including determination of the fundamental solution and a Cauchy integral formula. The study of the heat equation in this context seems to present interesting new ideas. To conclude, the problem to find the most general deformation of a Dirac operator still yielding an $\mathfrak{osp}(1|2)$ realization is expected to generate many new insights.

\section*{Acknowledgment}

 The author would like to thank F. Brackx, K. Coulembier, N. De Schepper and K. B. Wolf for their valuable comments with respect to this paper.

\end{document}